\documentclass[twocolumn]{aastex63}
\usepackage{amssymb, amsmath}
\usepackage{color}
\usepackage{morefloats}
\usepackage{hyperref}

\begin{document}

\title{The Red Supergiant Binary Fraction \\ as a Function of Metallicity in M31 and M33}
\email{kneugent@uw.edu}

\author[0000-0002-5787-138X]{Kathryn F.\ Neugent}
\affiliation{Department of Astronomy, University of Washington, Seattle, WA, 98195}
\affiliation{Lowell Observatory, 1400 W Mars Hill Road, Flagstaff, AZ 86001}


\begin{abstract}
Recent work measuring the binary fraction of evolved red supergiants (RSGs) in the Magellanic Clouds points to a value between 15-30\%, with the majority of the companions being un-evolved B-type stars as dictated by stellar evolution. Here I extend this research to the Local Group galaxies M31 and M33, and investigate the RSG binary fraction as a function of metallicity. Recent near-IR photometric surveys of M31 and M33 have lead to the identification of a complete sample of RSGs down to a limiting $\log L/L_{\odot} \geq 4.2$. To determine the binary fraction of these M31 and M33 RSGs, I used a combination of newly obtained spectroscopy to identify single RSGs and RSG+OB binaries as well as archival UV, visible and near-IR photometry to probabilistically classify RSGs as either single or binary based on their colors. I then adjusted the observed RSG+OB binary fraction to account for observational biases. The resulting RSG binary fraction in M33 shows a strong dependence on galactocentric distance with the inner regions having a much higher binary fraction ($41.2^{+12.0}_{-7.3}$\%) than the outer regions ($15.9^{+12.4}_{-1.9}$\%). Such a trend is not seen in M31; instead, the binary fraction in lightly reddened regions remains constant at $33.5^{+8.6}_{-5.0}$\%. I conclude the changing RSG binary fraction in M33 is due to a metallicity dependence with higher metallicity environments having higher RSG binary fractions. This dependence most likely stems not from changes in the physical properties of RSGs due to metallicity, but changes in the parent distribution of OB binaries.
\end{abstract}
 
\section{Introduction}
While the exact binary fraction of un-evolved massive OB stars is unknown, it is thought to be 50-60\% or higher (see discussion in \citealt{SanaSci}). As the stars in these OB binary systems burn through their hydrogen and progress towards departing the main sequence, several events can alter their evolutionary journeys. If the OB stars have close companions (with periods less than $\sim$1500 days), they will likely experience Roche lobe overflow (RLOF) as the two stars begin to interact and exchange material \citep{Moe2017, SanaSci}. When this occurs, either the mass transfer will be stable and create a stripped primary that never fully evolves (see \citealt{Kippenhahn1967, Yoon2017, Trevor2018}), or the mass transfer will be unstable, where the secondary accretes more mass than it can absorb. In this unstable case, the secondary then goes through RLOF and the two stars enter a common envelope phase. Depending on their proximity, the more massive star will continue to evolve past the OB phase before eventually merging with its less massive companion \citep{Chatzopoulos2020, Wheeler17}. However, in OB binary systems with larger separations (periods generally greater than $\sim$1500 days), the more massive star will evolve much like a single star -- after burning through its hydrogen, it will briefly pass through the yellow supergiant (YSG) phase before becoming a red supergiant (RSG). In this case, it will retain its less massive companion. Here we study these RSGs with less massive main-sequence companions and their prevalence in the Local Group galaxies M31 and M33 as a function of metallicity.

Stellar evolution dictates that the most common companions to RSG binaries are B-type stars (see extensive discussion in \citealt{NeugentRSG, NeugentRSGII, NeugentRSGIII}). To summarize, the least massive OB star that evolves into a RSG turns off the main sequence at around the same time a $3M_\odot$ star (A0V) enters the main sequence. Any star less massive than this will still be a protostar by the time the RSG has formed. Thus, from an evolutionary standpoint, we do not expect to find RSGs in binary systems with A,F,G,K, or M dwarfs. While it is possible for RSGs to have more massive companions than B-type stars (O stars, Wolf-Rayets, YSGs, etc.), the short lifetimes of these massive companions make such pairings rare, though such systems should ultimately lead to RSGs with compact object companions (see \citealt{Hinkle2020}). This is what we've seen observationally -- all known binary systems where the RSG was initially the most massive star have B-type companions (see Table 1 in \citealt{NeugentRSGII}). \citet{NeugentRSG} devised a set of photometric criteria to search for RSG+B star binaries based on the premise that RSG binaries would have excess flux in the blue compared to a single RSG. So far, over 250 new systems have been spectroscopically confirmed in the Small and Large Magellanic Clouds (SMC, LMC; described in \citealt{NeugentRSG, NeugentRSGIII} with some additional candidates mentioned in \citealt{GF2015}) and M31 and M33 (discussed first in \citealt{NeugentRSGII} with new discoveries described here).

\citet{NeugentRSGIII} went beyond simply identifying RSG binary systems to placing constraints on the overall binary {\it fraction} of RSGs in the LMC. We began by using near infrared (NIR) photometry from 2MASS \citep{2MASS} to first identify a complete sample of RSGs down to $\log L/L_\odot \geq 4$. We then spectroscopically confirmed a subset of the RSG+B star binary candidates before using a k-Nearest Neighbors (k-NN) algorithm to estimate the overall observed binary fraction. After taking observational biases into account, such as line-of-sight pairings and binaries in eclipse, as well as systems our photometric criteria were not designed to detect, we found a binary fraction of $19.5^{+7.6}_{-6.7}$\%. Here I expand this research to a wider range of metallicities by looking at M31 and M33.

While there is no doubt that the evolution of massive stars is drastically influenced by the surrounding environment's metallicity (see \citealt{Conti1975}), it is less clear whether metallicity influences the binary fraction of such stars, particularly for evolved massive stars. Many studies have been done on the binary fraction of Wolf-Rayet (WR) stars in the Local Group galaxies M31 and M33 \citep{WRbins} and the Magellanic Clouds \citep{Foellmi2003, FoellmiSMC, Bartzakos2001} and all research points to a consistent binary fraction of 30-40\% across all metallicities. But, no observational tests have looked at how the binary fraction of the less massive RSGs might change with metallicity. For RSG binaries, the separation between the two stars dictates whether a stable RSG binary system will form and any dependence of this parameter on metallicity has not been studied. Additionally, there is little evidence for a metallicity dependence in the binary fraction of un-evolved OB stars (e.g. \citealt{Moe2017, Moe2019}). Thus, it is not immediately clear whether the more evolved RSG binary fraction should be sensitive to metallicity.

To investigate the RSG binary fraction as a function of metallicity further, I've focused my efforts on the RSG content of M31 and M33 for several reasons. First, these two galaxies are close enough that we can observe a {\it complete} sample of RSGs down to a limiting luminosity. This is something we are unable to do in our own Milky Way because our location within a spiral arm makes it impossible to observe the majority of the Galaxy's RSGs given high extinction within the Galactic plane. Secondly, we've recently obtained NIR photometry of both M31 and M33 that allows us to identify a complete sample of RSGs and thus accurately compute the binary fraction \citep{M3133RSGs}. Finally, and most importantly, the combination of M31 and M33 allow us to target a large metallicity range from $\log(O/H)+12 = 8.3$ in the outer regions of M33 \citep{Magrini2007} to $\log(O/H)+12 = 8.9$ in M31 \citep{Sanders} (see a deeper discussion about the caveats behind these values in Section 5.3). This allows us to carefully investigate the effects of metallicity on the RSG binary fraction by just observing two galaxies. 

While a large potion of this paper focuses on the science related to identifying and confirming RSG binaries within M31 and M33, it would be impossible to determine the binary {\it fraction} without also understanding the total RSG content within each galaxy. Thankfully, \citet{M3133RSGs} recently used archival NIR photometry in M31 and M33 to identify a complete sample of RSGs down to a limiting luminosity of $\log L/L_{\odot} \geq 4.0$. Overall, they used a similar method to the one employed by \citet{NeugentRSGIII} when selecting LMC RSGs. They began by removing foreground stars using Gaia parallax and proper motion values \citep{Gaia} and then excluded asymptotic giant branch (AGB) stars due to their redder colors on the CMD and overall dimmer magnitudes. Finally, they transformed the NIR colors to effective temperature and luminosities before creating a final list of 7585 and 3911 RSGs in M31 and M33, respectively. This catalog serves as a starting point to calculate the binary fraction of RSGs in these two galaxies.

In Section 2, I describe our spectroscopic sample of confirmed single and binary RSGs in M31 and M33 before discussing how I used photometry and a k-NN approach to classify the remaining RSGs as either binary or single in Section 3. Section 4 focuses on the calculations behind determining the overall RSG binary fraction in M31 and M33 and leads into Section 5, which looks at the binary fraction as a function of metallicity. I then end with a brief summary and suggestions about possible next steps in Section 6.

\section{Spectroscopically Confirmed RSG+B Binaries}
\citet{NeugentRSGII} previously discovered 63 RSG+B star binaries in M31 and M33. Here I describe the selection of additional candidates, their followup spectroscopic confirmation, and the overall observed RSG binary sample used to determine the RSG+OB\footnote{So far, we've only detected {\it B-type} companions spectroscopically, but the method is also sensitive to RSG+O star binaries. However, due to the initial mass function, RSG+O star binaries should be much rarer as discussed above (also see discussion in \citealt{NeugentRSGIII}). Still, I have tried to use RSG+B when explicitly discussing the binaries we've observationally detected but RSG+OB when discussing the sensitivity of the method.} binary fraction. 

\subsection{Selection Criteria}
Overall, we expect RSG+OB binaries to have excess blue light originating from the OB star companion as compared to single RSGs. Thus, the spectral energy distribution (SED) of a binary RSG will appear flatter than that of a single RSG given the excess flux in the blue. This allows us to define a set of photometric color cuts that can be used to select RSG+OB binaries based off of both observed single RSG and OB-type stars as well as synthetic RSG+OB star binaries modeled using the MARCS RSG spectral models \citep{MARCS} in combination with the BSTARS06 spectral models \citep{BSTARS06}. \citet{NeugentRSG} used this information to define a set of color-color cuts in $U-B$ and $R-I$ that could be used for selecting RSG binaries (see their Figure 11). 

Ideally, candidate RSG+OB binaries would have been selected as a subset of the M31 and M33 RSGs from \citet{M3133RSGs} that had excess flux in the blue. This is the method \citet{NeugentRSGIII} used to select RSG binary candidates in the LMC. However, at the time of the 2019 spectroscopic observations (detailed below), the photometry described by \citet{M3133RSGs} was not available. Instead, I relied on photometry from a variety of photometric surveys to select RSG binary candidates. The methods used to select candidates therefore differed slightly based on the initial photometric catalog. In all cases, I set a limiting $V$ magnitude of 20 (it is not possible to observe fainter RSGs spectroscopically with sufficient S/N), and removed the crowded stars as dictated by the ``X" flag in the Local Group Galaxy Survey (LGGS, \citealt{LGGS}).

The initial candidate list for M31 and M33 was described in detail by \citet{NeugentRSG} and was selected only based on LGGS $U-B$ and $R-I$ photometry. It contained 138 potential binaries in M31 and 142 in M33. As discussed in \citet{NeugentRSG}, the assumption was that in this region of color-color space, contamination by other types of stars was small and all sources with these colors should be RSG+OB binaries. Follow-up observations of 149 of these candidates were done by \citet{NeugentRSGII} which lead to the discovery of 63 new RSG binary systems. They additionally identified a host of non-stellar contaminants such as QSOs and galaxies \citep{Massey2019}. However, the majority of the observed candidates were either RSGs, B-type stars or RSG+B type binaries. Thus, the remaining 131 candidates in M31 and M33 are included as candidates for spectroscopic confirmation.

Although the final selection of M33’s RSGs described in \citet{M3133RSGs} had not been completed, preliminary results were used to select stars for spectroscopic follow-up. There were two main differences between the preliminary and the final lists of RSG candidates. First, the preliminary cut took the UKIRT WFCAM source classification flags more stringently, insisting that all sources be described as ``stellar” on both the $J$ and $K_s$ list. Subsequent examination showed that these flags were not always consistent; i.e., the same object might be described as stellar on some but not all frames. Inspection of the frames revealed that this resulted in some valid sources being excluded in the preliminary list. Secondly, the position of the RSG region was refined in the color-magnitude diagram for the final list.  In the UKIRT photometry of a few regions in M31 as discussed in \citet{UKIRT}, the blue cut-off line for RSGs was defined as parallel to the red cut-off line, following the same procedure used by \citet{Yang2019} for selection of RSGs in their SMC catalog. However, as argued in \citet{M3133RSGs}, there is no physical basis for this. Instead, the blue cutoff has been redefined as a vertical line. Foreground stars were removed using Gaia results in much the same way as ultimately used for the final list, although some refinements were made \citep{M3133RSGs}. After cross-matching this list of M33 RSGs with the LGGS, I opted to keep all RSGs with $U-B < 1$, similar to what was done for the LMC binary candidates in \citet{NeugentRSGIII}. This resulted in some of the RSG binary candidates falling outside the region of $U-B$ vs.\ $R-I$ color-color space initially defined for candidate selection. But, with the addition of NIR photometry, I was curious to learn more about the nature of these RSGs with excess blue flux. After these photometric cuts, 182 candidate RSG binaries remained in M33. Of these stars, 134 of them were new additions and the remaining 48 were duplicates from the candidates described by \citet{NeugentRSG}. Combined with the 63 additional candidates from \citet{NeugentRSG}, I identified 245 RSG binary candidates in M33. 

Selecting RSGs in M31 was a bit more complicated as I did not have access to the NIR photometry as described in \citet{M3133RSGs}. Instead, I relied on a combination of the RSGs identified using NIR photometry in a small portion of M31 as described by \citet{UKIRT} and the 2MASS 6$\times$ ``long exposure" M31 photometry which goes around 1 magnitude deeper than the standard 2MASS survey \citep{2MASS}. Both of these lists were cross-matched with Gaia to remove foreground stars using the method described in \citet{UKIRT}. Again, they were cross-matched with the LGGS and a cut was made for stars with $U-B < 1$. In total, we identified 214 new RSG binary candidates in M31. When combined with the remaining 36 candidates from \citet{NeugentRSG}, I identified 250 RSG binary candidates in M31. 

\begin{figure}
\includegraphics[width=0.5\textwidth]{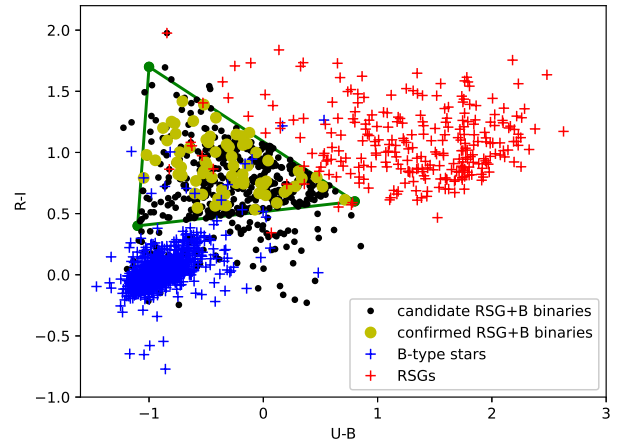}
\caption{\label{fig:colorCutCands} $U-B$ vs.\ $R-I$ color-color plot of candidate RSG binaries. The blue and red $+$s represent the LGGS photometry from spectroscopically confirmed B-type stars and RSGs, respectively. The yellow dots are the spectroscopically confirmed RSG+B star binaries discussed in \citet{NeugentRSGII}. The green lines represent the region in $U-B$ and $R-I$ space that we based our initial candidate list off of using only the LGGS photometry. The vertices of the three green points in $U-B$ vs.\ $R-I$ color space are (-1.1,0.4), (0.8,0.6), and (-1,1.7). Finally, our candidate M31 and M33 RSG+B star binaries are shown as black points. For stars with NIR photometry that could be used to classify them as RSGs, we allowed all values of $R-I$ but restricted the candidates to those with $U-B < 1$. This is why there are some candidates that fall outside of the green triangle.}
\end{figure}

The location of the 245 M33 candidates and 250 M31 candidates within $U-B$ vs.\ $R-I$ color space are shown in Figure~\ref{fig:colorCutCands}. Additionally plotted are the known RSG binaries, single RSGs and single B-type stars previously spectroscopically confirmed and described in \citet{NeugentRSG, NeugentRSGII}.

\subsection{Observations and Reductions}
Observations were carried out using the multi-object, fiber-fed spectrograph Hectospec \citep{Fabricant2005} on the 6.5-m MMT located near Tucson, AZ. Collaborator P. Massey and I were assigned 1.5 nights of queue-scheduled dark time in Fall 2019 through the Arizona Time Allocation Committee. However, due to poor weather and an instrument failure, only a single M31 and a single M33 configuration were observed (one third of what we were hoping to obtain). The M31 field was observed on UT 2019 September 5 and the M33 field was observed on UT 2019 October 22 through clear sky conditions with exposure times of 3 $\times$ 3000 seconds. The data were taken with the 270 line mm$^{-1}$ grating that yields a spectral coverage from 3700 - 9000 \AA. Each of the 300 fibers is 250 $\mu$m which provides a spectral resolution of 6\AA. Reductions were carried out using v2.0 of the Hectospec REDuction (HSRED) code. 

\subsection{Results}
As part of the Fall 2019 observing run, 203 of the 495 new RSG binary candidates were spectroscopically observed in M31 and M33. Of these, 124 (61\%) of them turned out to be RSG+B star binaries (60 in M31 and 64 in M33), 38 (19\%) are single RSGs (21 in M31 and 18 in M33), 20 (10\%) only show upper Balmer lines (8 in M31, 12 in M33), and the remaining 21 (10\%) were other accidental discoveries (3 QSOs, 7 HII regions, and the remaining with poor S/N). Combining this with the sample discussed in \citet{NeugentRSGII}, this brings the total number of RSG+B star binaries known in M31 up to 88 and in M33 to 94. Further information about these spectroscopically observed stars, their LGGS photometry, and classifications, can be found in Table~\ref{tab:specClass}.

\begin{deluxetable*}{l l l l r r r r l}
\tabletypesize{\scriptsize}
\tablecaption{\label{tab:specClass} Spectroscopically Observed M31 and M33 Stars\tablenotemark{\scriptsize*}}
\tablewidth{0pt}
\tablehead{
\colhead{LGGS Designation\tablenotemark{\scriptsize a}}
&\colhead{$\alpha_{2000}$}
&\colhead{$\delta_{2000}$}
&\colhead{$V$}
&\colhead{$B-V$}
&\colhead{$U-B$}
&\colhead{$V-R$}
&\colhead{$R-I$}
&\colhead{Classification}
}
\startdata
J003744.37+400401.5 & 00:37:44.36 & +40:04:01.4 &   19.783 &  1.528 &  0.384 &  0.675 &  0.578 & RSG \\
J003848.43+400432.4 & 00:38:48.42 & +40:04:32.3 &   19.169 &  1.571 &  0.380 &  0.481 & -0.229 & RSG \\
J004059.67+402049.6 & 00:40:59.66 & +40:20:49.5 &   18.881 &  1.343 &  0.484 &  0.642 &  0.738 & RSG \\
J004156.45+402004.5 & 00:41:56.44 & +40:20:04.4 &   19.145 &  1.587 &  0.217 &  0.609 &  0.361 & RSG \\
J004239.02+413643.2 & 00:42:39.01 & +41:36:43.1 &   19.667 &  0.482 &  0.289 &  0.275 &  0.572 & RSG+B \\
J004248.13+414714.5 & 00:42:48.12 & +41:47:14.3 &   18.895 &  1.193 & -0.367 &  1.091 &  1.177 & RSG+B \\
J004248.94+414035.3 & 00:42:48.93 & +41:40:35.1 &   19.532 &  0.966 & -0.710 &  0.848 &  0.884 & RSG \\
J004306.93+413808.1 & 00:43:06.92 & +41:38:08.0 &   19.748 &  0.847 & -0.228 &  0.535 &  0.545 & OB \\
J004307.61+415103.8 & 00:43:07.60 & +41:51:03.6 &   19.705 &  1.083 & -0.489 &  0.823 &  0.798 & RSG+B \\
J004307.87+413407.6 & 00:43:07.86 & +41:34:07.5 &   19.196 &  1.038 &  0.415 &  0.728 &  0.690 & RSG \\
J004315.34+414242.0 & 00:43:15.33 & +41:42:41.8 &   19.966 &  0.276 & -0.221 &  0.371 &  0.704 & RSG+B \\
J004319.73+414033.3 & 00:43:19.72 & +41:40:33.1 &   18.725 &  0.899 & -0.115 &  0.523 &  0.546 & OB \\
J004332.51+415232.0 & 00:43:32.50 & +41:52:31.8 &   19.747 &  0.809 & -0.425 &  0.711 &  0.766 & RSG+B \\
J004334.40+414915.9 & 00:43:34.39 & +41:49:15.7 &   20.243 &  0.841 & -0.485 &  0.872 &  0.975 & RSG+B \\
J004340.58+412550.0 & 00:43:40.57 & +41:25:49.9 &   18.952 &  0.513 & -0.322 &  0.476 &  0.525 & RSG+B \\
\enddata
\tablenotetext{*}{This table is published in its entirety in the machine-readable format. A portion is shown here for guidance regarding its form and content.}
\tablenotetext{a}{All photometry is from the LGGS; \citealt{LGGS}}
\end{deluxetable*}

\section{Classifying RSGs as Single or Binary}
Thanks to the recent work done by \citet{M3133RSGs}, the RSG content of both M31 and M33 is now well defined. Here I describe how I used a k-NN algorithm to classify RSGs that were not observed spectroscopically as either single or binary following the procedure initially outlined in \citet{NeugentRSGIII}. 

\subsection{The Initial Sample}
The initial set of RSGs was selected photometrically based on NIR colors as described by \citet{M3133RSGs}. While the overall process of removing foreground stars using Gaia, and then defining appropriate $J$ and $J-K$ color cuts was the same for both galaxies, the resulting datasets have their own slight differences and caveats. For this reason, I'll discuss both M31 and M33 separately below. One prominent difference between the initial sample used here for M31 and M33 vs.\ the initial sample used by \citet{NeugentRSGIII} for the LMC is that the limiting luminosity has been changed from $\log L/L_\odot \geq 4.0$ to $\log L/L_\odot \geq 4.2$. This is primarily due to increased crowding in the inner regions of M31 (see further discussion in \citealt{M3133RSGs}). To keep the comparisons consistent across galaxies, I've thus adjusted all of the RSG lower luminosity completeness limits to $\log L/L_\odot \geq 4.2$, corresponding to a $\sim10M_\odot$ initial solar mass star. 

\subsubsection{M33}
Starting with the list of M33 RSGs presented by \citet{M3133RSGs} and selecting those with $\log L/L_\odot \geq 4.2$, yielded 1970 RSGs. Since I later rely on LGGS photometry for the k-NN classification, I additionally selected M33 RSGs that were within the LGGS survey area. This limited me to the 1702 RSGs with $\rho < 1$ where $\rho$, the galactocentric distance, assumes a Holmberg radius of 30.8 arcminutes, an inclination of 56.0$^\circ$, and a position angle of the major axis of 23.0$^\circ$. At a distance of 830 kpc \citep{vandenbergh2000}, $\rho = 1$ corresponds to 7.44 kpc. These 1702 stars formed my initial list of M33 RSGs to classify as either single or binary. 

After spectroscopically confirming 94 total RSG+B star binaries in M33, I was now left with 65 after removing 12 binaries with $\log L/L_\odot < 4.2$, 5 with $\rho > 1$, 6 that Gaia suggests as possible foreground stars, and 6 that are a bit too yellow to be classified as RSGs. Using spectra from \citet{DroutM33RSGs, Massey98}, and our own Hectospec data described above and in \citet{NeugentRSGII}, I additionally classified 230 RSGs as single. Thus, 17\% of our initial M33 RSGs have been spectroscopically confirmed as either binary or single. 

Of these 1702 RSGs, 80\% have LGGS counterparts and thus $B$, $V$, and $R$ photometry. The remaining 218 stars are primarily in crowded regions where the LGGS was not able to accurately distinguish individual stars (see \citealt{M3133RSGs} for more explanation). However, as described below, I was able to supplement the LGGS dataset with additional photometry to help classify many of the stars without LGGS data. Of the 1364 stars with LGGS photometry, 996 (73\%) additionally have $U$ photometry which greatly helps when classifying the stars as binary vs.\ single due to the excess blue flux.

\subsubsection{M31}
I followed a similar process for M31, but soon realized that the effects of crowding were much more pronounced in M31 than they were in M33 or the LMC. To reduce issues with crowding, I removed stars in the innermost portion of M31 with $\rho < 0.1$. I additionally limited stars to those with $\rho <= 0.75$ to constrain the survey area to approximately what was covered by the LGGS. In M31, $\rho$ was computed assuming a Holmberg radius of 95.3 arcminuntes, an inclination of 77.0$^\circ$, and a position angle of the major axis of 35.0$^\circ$. At a distance of 760 kpc \citep{vandenbergh2000}, $\rho = 1$ corresponds to 21.07 kpc. After making the luminosity and $\rho$ cuts, I was left with 1909 RSGs in M31. 

Of the 88 spectroscopically confirmed RSG+B star binaries in M31, only 43 were in this subset. As with M33, the removals were due to $\log L/L_\odot < 4.2$ (11 stars), $\rho$ out of limits (15 stars), foreground classification (5 stars), and outside RSG color limits (14 stars). The large number of spectroscopically confirmed binaries outside of the color range is due to excess reddening in M31, as discussed later. I additionally suspect the stars removed as foreground stars due to Gaia classifications are likely members based on Ca\,{\sc ii} triplet radial velocity measurements. However, these stars have still been removed from the binary fraction calculation for consistency's sake. An additional 183 spectroscopically confirmed single RSGs from \citet{Massey98, MasseyEvans16}, the Hectospec data described above and in \citet{NeugentRSGII} brought the total number of M31 RSGs with spectra to 226, or 12\%. 

Crowding and extinction issues in M31 also decreased the number of RSGs with LGGS counterparts. Out of the 1909 RSGs, only 76\% have LGGS $B$, $V$, and $R$ photometry. As I discuss later, high reddening in certain parts of M31 may lead to additional issues, and thus I later calculate the binary fraction of RSGs in M31 for both the entire galaxy and just in the less reddened regions.

\subsection{UV Photometry}
RSG+OB binaries should have strong flux in the UV compared to single RSGs which have relatively little signal at such short wavelengths. Thus, similar to what \citet{NeugentRSGIII} did with the UV Galaxy Evolution Explorer (GALEX; \citealt{Simons14, Martin05, Morrissey07}) data in the LMC, I cross-matched the M31 and M33 RSGs with photometry from the Panchromatic Hubble Andromeda Treasury (PHAT) dataset \citep{PHAT} and the soon-to-be-published Panchromatic Hubble Andromeda Treasury: Triangulum Extended Region (PHATTER) dataset (Williams et al., in prep). Due to the vastly increased spatial resolution of the PHAT and PHATTER data compared to our ground-based NIR photometry, cross-matching was slightly more complicated than just taking the closest match in coordinates. Instead, I searched for the closest star within $0\farcs5$ that had a similar F110W magnitude to the RSG's $J$ magnitude. While the HST F110W and 2MASS $J$ bandpasses aren't identical, this method proved quite successful based off a visual check of a random set of around 100 RSGs.

By including both the PHAT and PHATTER data, I was able to add UV photometry to a set of stars that previously only had LGGS visible and NIR photometry. There were also a small number of RSGs without LGGS photometry that instead had PHAT or PHATTER photometry and thus could be classified as a single or binary RSG. In M31, 314 of our RSGs had matches in PHAT with 72 of them having no LGGS match. In M33, 696 had matches in PHATTER with 60 of them having no LGGS match. The number of stars with photometry from each survey is shown in Table~\ref{tab:kNN}.

\subsection{k-NN Classification}
To calculate the observed RSG+OB star binary fraction, I relied on a k-NN approach, much as was done by \citet{NeugentRSGIII} for the LMC RSGs. The rational behind this approach is that RSGs with OB star companions should appear photometrically different from single RSGs due to their increased flux in the blue. Using our spectroscopically confirmed RSG binaries and single stars and their corresponding photometry as inputs, the k-NN algorithm assigns a probability of binarity to each of the remaining RSGs. RSGs with colors similar to spectroscopically confirmed RSG binaries are assigned a higher probability of being a binary than RSGs with colors similar to spectroscopically confirmed single RSGs. 

The k-NN algorithm was implemented using Python's {\sc scikit-learn} machine-learning package. I trained and tested the data using k-fold cross validation of the spectroscopically confirmed stars mentioned above after scaling the data using {\sc scikit-learn}'s RobustScaler. For both M31 and M33, there were essentially three different datasets to classify: those with just LGGS photometry, those with LGGS and HST (either PHAT or PHATTER) photometry, and those with just HST photometry. Then there were the remaining stars with neither LGGS or HST photometry that could not be classified using the k-NN algorithm. The input parameters for the k-NN algorithm (including the number of folds and neighbors used), star counts, and the accuracy rates are shown in Table~\ref{tab:kNN}. 

\begin{deluxetable*}{l l l l l l l l l}
\tabletypesize{\scriptsize}
\tablecaption{\label{tab:kNN} k-NN Parameters}
\tablewidth{0pt}
\tablehead{
\colhead{Galaxy}
&\colhead{LGGS}
&\colhead{HST}
&\colhead{\# binary RSGs}
&\colhead{\# single RSGs}
&\colhead{\# unknown RSGs}
&\colhead{\# folds}
&\colhead{\# neighbors}
&\colhead{Accuracy \%}
}
\startdata
M31 & $\checkmark$ & $\checkmark$ & 43 & 186 & 209 & 4 & 10 & 91.0 \\
M31 & $\checkmark$ &  & 43 & 186 & 1009 & 4 & 13 & 98.7 \\
M31 &  & $\checkmark$ & 9 & 24 & 72 & 4 & 8 & 76.4 \\
M31 &  &  & \nodata & \nodata & 390 & \nodata & \nodata & \nodata \\ \hline
M33 & $\checkmark$ & $\checkmark$ & 65 & 230 & 593 & 4 & 15 & 94.3 \\
M33 & $\checkmark$ &  & 65 & 230 & 708 & 7 & 10 & 94.3 \\
M33 &  & $\checkmark$ & 18 & 25 & 60 & 6 & 10 & 80.0 \\
M33 &  &  & \nodata & \nodata & 46 & \nodata & \nodata & \nodata \\
\enddata
\end{deluxetable*}

\begin{figure*}
\includegraphics[width=0.5\textwidth]{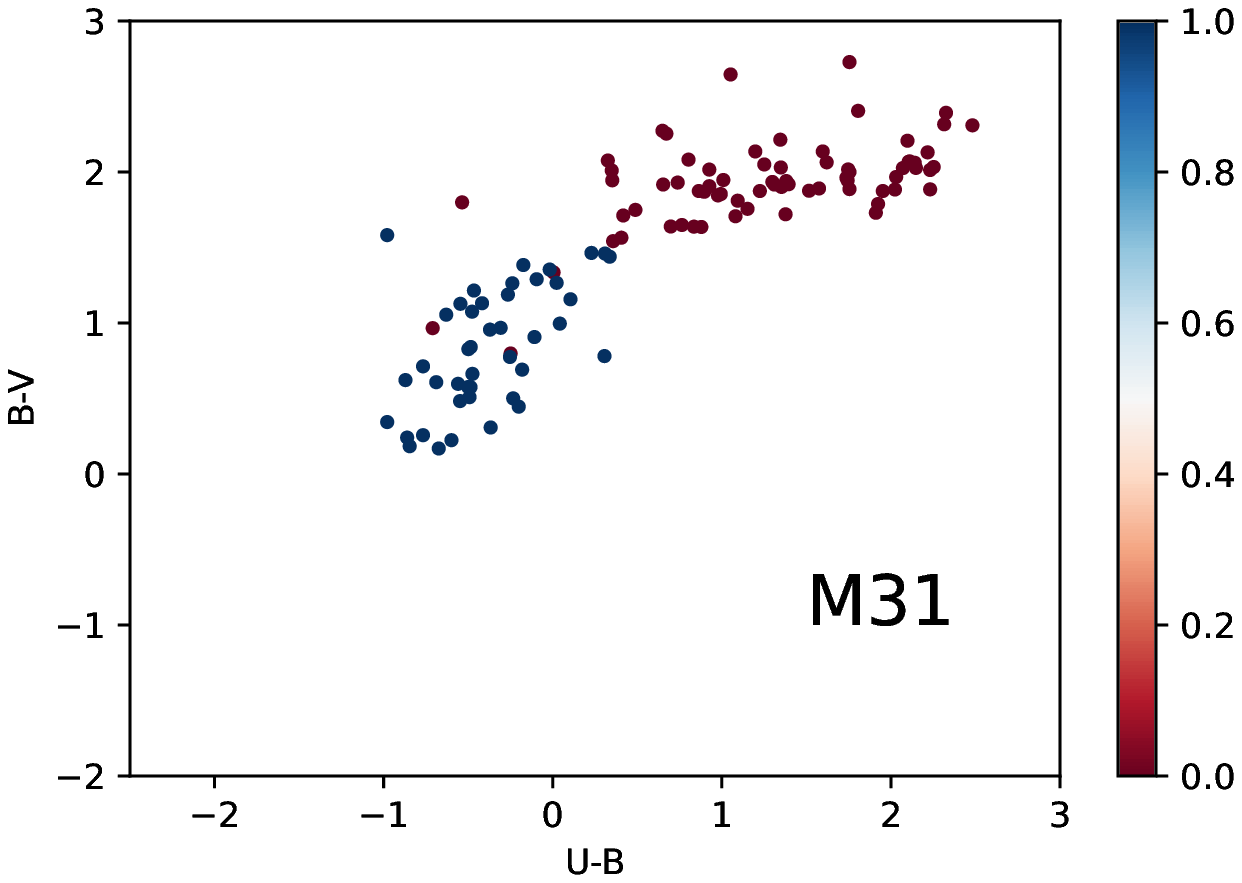}
\includegraphics[width=0.5\textwidth]{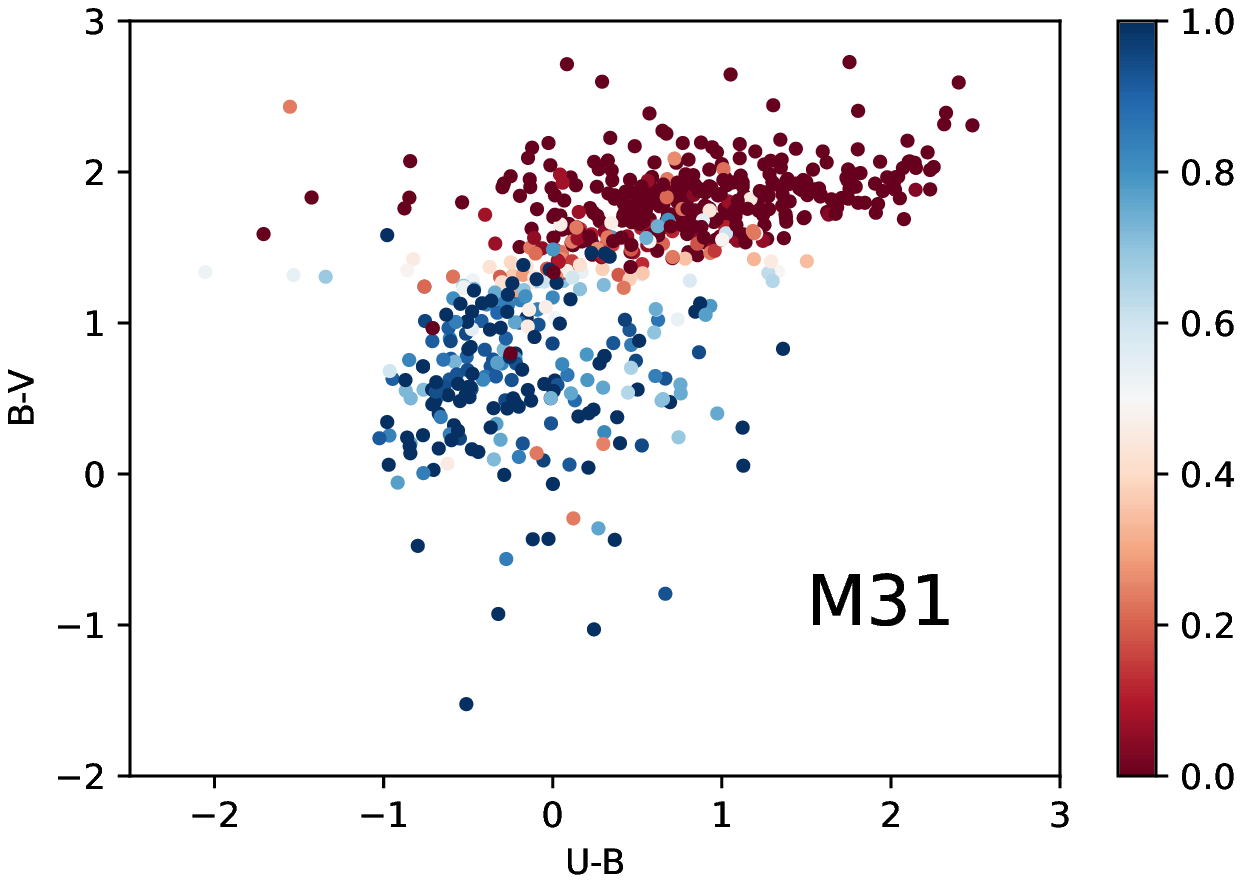}
\includegraphics[width=0.5\textwidth]{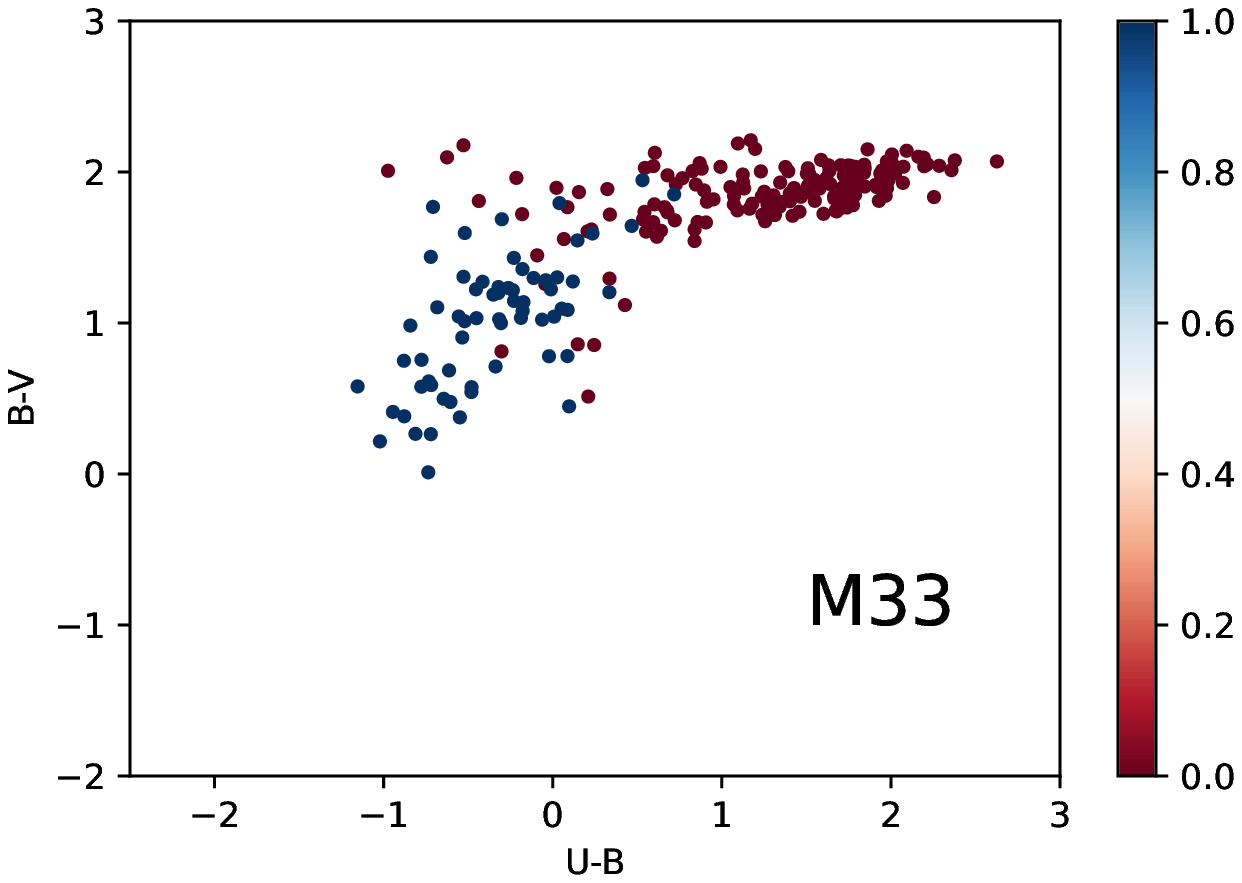}
\includegraphics[width=0.5\textwidth]{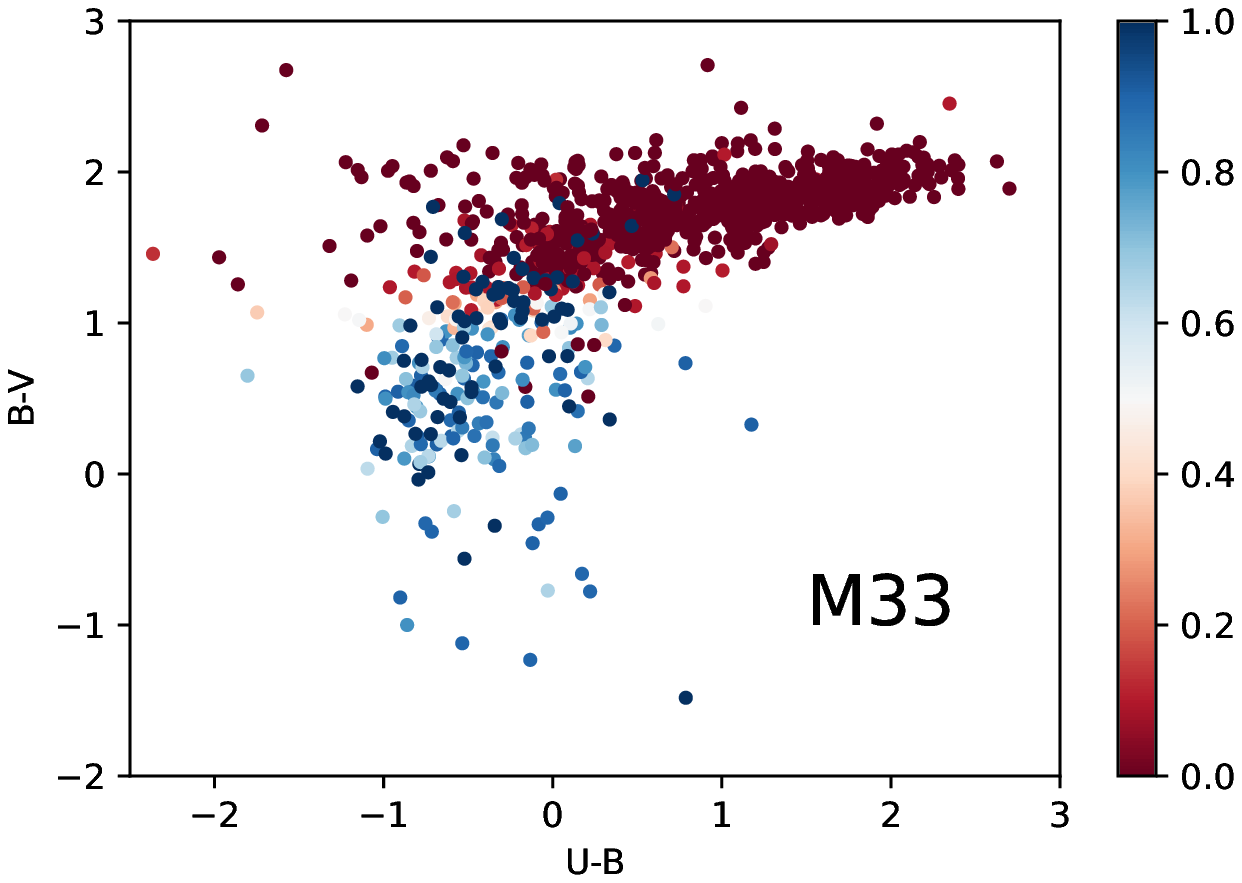}
\caption{\label{fig:kNN} Results of k-NN classification for M31 and M33. Figures on left show the locations in $U-B$ vs.\ $B-V$ color space of the spectroscopically confirmed RSG+B binaries and single stars in M31 (top) and M33 (bottom). Figures on the right show the k-NN classified RSGs as colored by their probability of binarity with bluer colors representing a higher probability of having a B-type binary companion and redder colors representing a higher probability of being a single RSG. As expected, RSG+B binaries have lower $U-B$ colors coming from the B-type star. Note the surplus of M31 RSGs in the color regime where both $U-B$ and $B-V$ is between 0 and 1 (top, right), and lack of such stars in the remaining three plots. As discussed in the paper and in \citet{M3133RSGs}, we believe these stars to be more highly reddened based on their location and thus their colors to be anomalous.}
\end{figure*}

A visual representation of the k-NN classification results is shown in Figure~\ref{fig:kNN} where the individual points are color-coded based off their likelihood of having an OB-type companion (bluer points correspond to higher likelihood). The two figures on the left show the $U-B$ and $B-V$ colors of the initial spectroscopically confirmed RSG+B binaries and single RSGs for both M31 and M33. As expected, the RSG+B binaries show excess blue flux coming from the B-type companion and thus their $U-B$ colors are more negative. The two figures on the right show the results of the k-NN classification algorithm. Note that, as with the LMC dataset described by \citet{NeugentRSGIII}, the ``transition" period between a RSG having an OB-type companion is around a $U-B = 0$. RSGs with $U-B$ values higher than this are almost always single RSGs and those with $U-B$ values lower than this likely have an OB-type companion. An additional constraint can be made with $B-V$ values as RSGs with $B-V > 1$ appear much more likely to be single RSGs. An interesting aspect of the M31 plot to note is the cluster of classified RSGs with both $U-B$ and $B-V$ values between 0 and 1 (see upper right plot). Stars with these colors were not observed spectroscopically (see upper left plot) and yet a large number of them exist. This highlights the effect highly reddened regions can have on the photometry and the need to be careful when selecting stars to use to determine the overall observed binary fraction. 

\subsection{Completeness Issues}
As with any observational survey, there are completeness issues that must be mentioned. For this particular survey, there are two primary concerns. The first is that the survey didn't detect the correct total number of RSGs and the second is that the survey didn't detect the correct number of binaries. Here I discuss both concerns.

For this paper, I am concerned with the binary {\it fraction} of RSGs in M31 and M33. Thus, it is acceptable for the survey to be incomplete in some respects as long as it is incomplete equally for both the single and the binary RSGs. This is why, for example, I've made a cut in luminosity at $\log L/L_{\odot} \geq 4.2$. While, as detailed in \citet{M3133RSGs}, RSGs with luminosities below this limit have been detected and spectroscopically confirmed, a {\it complete sample} for lower luminosity RSGs is not known due primarily to crowding and reddening. Another example deals with the cuts in $\rho$ for both M31 and M33. As discussed earlier, I removed the inner bulge of M31 ($\rho < 0.1$) from the sample set due to crowding. While I was able to identify RSGs within M31's bulge, it was increasingly difficult to match them to their LGGS counterparts. However, I do not expect crowding to change the binary fraction of RSGs because it should equally effect both the binary and single stars.

The more serious concern is that the survey isn't sensitive to the faintest OB-type companions. When searching for RSG+OB binaries in the LMC, \citet{NeugentRSGIII} were able to simply increase the exposure times in real time in order to achieve the S/N needed ($\sim100$ per 2\AA\ at 4000\AA\ as determined by \citealt{NeugentRSG}) to confidently rule out the presence of a faint OB-type companion. However, the Hectospec data taken in M31 and M33 were obtained in ``queue" mode and exposure times could not be adjusted in real time. Additionally, while Hectospec's multi-fiber capabilities make it an ideal instrument for obtaining spectra of hundreds of stars at a time, it comes at the price of throughput, particularly in the blue where the upper Balmer lines reside. Finally, Hectospec has a spectral resolution of 5\AA\ which is not small enough to detect the weakest lined B companions. Thus, the average S/N per 5\AA\ spectral resolution element at 4000\AA\ was $\sim$25 with a few spectra as low as 10 (though many as high as 40). 

To determine the type of B companion that could still be hiding in the spectrum of our visually single RSGs, I turned back to the modeling work done by \citet{NeugentRSG}. We created a large set of synthetic spectra of RSG+B binaries by combining BSTAR06 \citep{BSTARS06} models with MARCS RSG models \citep{MARCS} at various temperature and luminosity regimes. I took these models and degraded the resulting synthetic spectra to a S/N of 25 per Hectospec spectral resolution element. I then looked through the results to determine which spectra appeared to be single even though a binary companion was present. I was relieved to find that all of the B-type giant and supergiants companions would still be detectable along with a sizable number of B dwarfs. The limiting spectral type was a B3~V with a temperature of 20,000K. With a S/N of 25, we cannot detect RSG+B star binaries where the B-type companion is fainter than a B3~V. However, as I discuss later on in Section 4.4, it is possible to quantify this completeness limit and take it into account when determining the errors. 

We have a similar issue when it comes to the HST PHAT/PHATTER photometry in that the far-UV F275W photometry does not go deep enough to detect the faintest B dwarfs. This means that the k-NN algorithm might assign a lower probability of binarity to a RSG than it should. To investigate the extent of this issue, I considered two aspects. First, unlike with the GALEX photometry in the LMC where \citet{NeugentRSGIII} only had one UV datapoint, the HST photometry provided photometric coverage across a wide wavelength range with two measurements in the UV, two in the Visible and two in the NIR. Thus, even if the bluest filter could not detect the presence of the faintest B dwarfs, the blue excess would still be detectable in the other UV filter (F336W) and the two Visible filters. Furthermore, I found that, given the limiting magnitude of the PHAT and PHATTER surveys in the F275W filter of 25th magnitude \citep{PHAT}, the surveys are sensitive down to a B5~V in either M31 or M33, or a B-type V with a temperature of 15,400K \citep{AAQ}. However, after doing the same calculation for the F336W filter, even the faintest B dwarfs would be detectable with the M31 and M33 PHAT/PHATTER datasets. Thus, this does not add any completeness issues to the survey.

\section{The Overall Binary Fraction}
While the observations and surveys discussed above can provide an observed binary fraction, there are several observational biases that must be considered before a final binary fraction can be calculated for M31 and M33. The first set deals with biases present within the observations such as eclipsing binaries and line-of-sight pairings, while the second deals with the RSG binaries that the observations would not have been able to detect. To determine both the percentage of eclipsing binaries, and companions the survey was not sensitive to, I relied on the Binary Population and Spectral Synthesis (BPASS) models (v2.2.1) from \citet{BPASS1, BPASS2}. BPASS uses (among other initial conditions) the recent results of \citet{Moe2017} to create a population of single and binary stars in various metallicity environments. At a given time step, it provides values for periods, mass-ratios, separations (of binaries) and a variety of physical properties such as temperature and luminosity (for all stars). For the work done here, I was able to select the BPASS models with temperature and luminosity criteria matching the RSG cutoffs used here and described by \citet{M3133RSGs} to form the M31 and M33 RSG catalog. I additionally used metallicity values appropriate to the various environments of M31 and M33. Final values for the binary fraction in the various environments of M31 and M33 are given in Table~\ref{tab:results}.

\begin{deluxetable*}{l l l l l l l l l}
\tabletypesize{\scriptsize}
\tablecaption{\label{tab:results} Binary Fraction}
\tablewidth{0pt}
\tablehead{
\colhead{$\log(O/H)+12$\tablenotemark{\scriptsize *}}
&\colhead{BPASS ($Z$)}
&\colhead{Galaxy Region}
&\colhead{RSG+OB (\%)}
&\colhead{eclipsing (\%)}
&\colhead{line-of-sight (\%)}
&\colhead{RSGs+CC (\%)}
&\colhead{Total (\%)}
}
\startdata
8.29 & 0.004 & M33 (outer) & $7.9^{+10.8}_{-1.7}$ & $1.84 \pm 0.01$ & $1.6\pm0.8$ & $4.73 \pm 0.01$ & $15.9^{+12.4}_{-1.9}$\\
8.41 & 0.006 & M33 (middle) & $15.2^{+6.1}_{-5.4}$ & $1.84 \pm 0.01$ & $1.6\pm0.8$ & $4.73 \pm 0.01$ & $26.9^{+8.9}_{-5.5}$\\
8.72 & 0.008 & M33 (inner) & $27.6^{+7.8}_{-7.3}$ & $1.84 \pm 0.01$ &$1.6\pm0.8$ & $4.73 \pm 0.01$ & $41.2^{+12.0}_{-7.3}$\\ \hline
8.93 & 0.014 & M31 (all) & $20.4^{+6.8}_{-6.3}$ & $1.84 \pm 0.01$ & $1.2\pm0.6$ & $4.73 \pm 0.01$ & $32.7^{+10.1}_{-6.3}$\\
8.93 & 0.014 & M31 (lightly reddened) & $22.5^{+5.2}_{-5.0}$ & $1.84 \pm 0.01$ & $1.2\pm0.6$ & $4.73 \pm 0.01$ & $33.5^{+8.6}_{-5.0}$\\
\enddata
\tablenotetext{*}{M31 oxygen abundance from \citealt{Sanders}; M33 oxygen abundances from \citealt{Magrini2007}}
\end{deluxetable*}

\subsection{Eclipsing Binaries}
The majority of the observed binary fraction estimate is based on single epoch photometry from LGGS or PHAT/PHATTER and thus it is important to consider the scenario when the OB companion is behind the RSG at the time of observation. An example system is VV Cep, the Galactic RSG+B star binary system with a 20.3 year orbit and 18 month secondary eclipse \citep{Bauer2000}. A similar system in M31 or M33 would not be identified as a potential binary during secondary eclipse because the excess flux from the OB companion would be masked. To determine the likelihood of such a positioning, I turned to BPASS. By examining the BPASS models at a single point in time (corresponding to single-epoch photometry), I used a script kindly provided by J.J.\ Eldridge to determine the maximum angle of inclination for eclipses based on the RSG's radius and separation and then further determined the likelihood of each system being in eclipse. Overall, the percentage of RSG+OB binaries in eclipse at any given time is $1.84\pm0.01$\%, and is given in Table~\ref{tab:results}.

\subsection{Line-of-Sight Pairings}
Line-of-sight pairings are increasingly a concern as we move to RSGs in more distant galaxies like M31 and M33. Such pairings exist when the RSG and OB star are both members of M31 or M33 but are not gravitationally bound and instead just exist on the same sight path. To estimate the number of spectroscopically confirmed RSG+B binaries that might be line-of-sight pairings instead of gravitationally bound companions, I used a Monte Carlo simulation much like the one done by \citet{NeugentRSGIII} to determine the likelihood of line-of-sight pairings for the RSGs based on their local OB star densities. Overall, for each of the spectroscopically confirmed RSG+B star binaries, I used $B$ and $V$ photometry and coordinate information from the LGGS to determine the positions of OB stars that fell within a $5\arcmin$ radius of the binary. I then randomly placed the binary within this region and checked to see if it fell within $0\farcs75$ (the radius of a MMT Hectospec fiber) within one of the OB stars. If it did, it was flagged as a line-of-sight pairing. OB stars were selected based on having $(B-V) < 0.0$ corresponding to an A0V, $V$ brighter than 22nd, and average $E(B-V)$ reddenings of M31 and M33 of 0.13 and 0.12, respectively \citep{LGGSIII}. Overall, I found a $1.2\pm0.6$\% chance of a line-of-sight pairing in M31 and a $1.6\pm0.8$\% chance in M33.

Adding in photometry will only make the situation improve. A large fraction of the RSGs (both single and binary) have HST photometry from PHAT / PHATTER (see Table~\ref{tab:kNN}) which provided images (and photometric catalogs) with milli-arcsecond resolution (see \citealt{PHAT} and Williams et al., in prep for details). Thus, the k-NN classification will not produce errors as high as those produced by spectroscopic confirmation. However, I still adopt the higher line-of-sight error estimates for the entire sample as a ``worst-case" scenario. Line-of-sight pairings are additionally discussed further in Section 5.1.

\subsection{RSGs and non-OB Companions}
The observational study outlined here is sensitive to RSGs with OB companions, which stellar evolution dictates are the most likely companions. However, it is important to consider the likelihood of other types of companions to RSGs. For stellar companions, this list includes evolved massive stars such as YSGs, other RSGs, and WRs. For non-stellar companions, this includes compact companions such as neutron stars and black holes.

As discussed extensively in \citet{NeugentRSGIII}, we do not expect to find a large number of RSG+RSG or RSG+YSG systems simply due to the short lifetimes of stars in these phases (the YSG phase, for example, only lasts tens of thousands of years). However, it is important to point out that the study described here would not be sensitive to such systems. Such a pairing might be observable through high resolution spectroscopy of the Ca\,{\sc ii} triplet line and searching for changes over time but this is outside the scope of the current work. Extensive efforts have recently been done by \citet{Patrick2019} and \citet{Dorda2020} searching for RSG binaries using radial velocity variations (and is discussed more below) and it is possible that this path will allow for the identification of any RSG+RSG/YSG binary systems through further analysis and orbit determinations.

Turning to more evolved massive stars, there cannot be a large population of missing RSG+WR systems. The WR content of both M31 and M33 is thought to be complete \citep{M31WRs, M33WRs} and the optical spectra of the known WRs have been examined for the presence of TiO bands indicative of a RSG companion. One star, J004453.06+412601.7, was originally classified by \citet{M31WRs} as a WN+TiO, but PHAT imaging clearly shows the TiO as coming from an object other than the WR. Thus, there are no known RSG+WR systems expected to be discovered in M31 or M33.

While RSG stellar companions should primarily be OB stars, we do expect a small fraction of RSGs to have compact companions that are not detectable using the observational method described here. Such compact companion systems occur when the RSG is the initially less massive star in the system and the more massive star explodes as a supernova and leaves behind either a neutron star or black hole as the companion to the RSG. The first spectroscopic confirmation of such a system was recently presented by \citet{Hinkle2020} for Galactic X-ray binary 4U 1954+31. They found that the system contains a high mass M-type supergiant paired with a neutron star and present a promising first detection for RSGs with compact companions. Additional, albeit circumspect, evidence of possible RSGs + compact companions was presented by \citet{Dorda2020} and their radial velocity study of RSGs in the LMC. Three of the LMC RSGs they classified as binaries were additionally spectroscopically observed by \citet{NeugentRSGIII} and show no evidence of an OB companion. It is thus possible these RSGs could be the first extragalactic detection of RSGs with compact companions. However, either further spectroscopy (both for radial velocities and companion detection) or a targeted x-ray campaign is necessary. Using BPASS v2.2.1, it is possible to estimate the fraction of RSGs with compact companions as $4.73\pm0.01$\%, as is shown in Table~\ref{tab:results}.

\subsection{RSGs and Faint B-type Companions}
As discussed in Section 3.4, the MMT spectra were not sensitive to B-type companions fainter than B3Vs because of S/N limitations. However, with BPASS v2.2.1, it is possible to determine the percent of RSG binaries with B-type companion less massive than a B3V (or $T_{\rm eff} < 20000$K). Using the same set of BPASS models described above, I was able to determine that at a given metallicity, 10\% of RSG binaries will contain a B-type companion that is fainter than a B3V. Thus, for any given binary fraction, the errors in the direction of {\it increasing} the binary fraction have been made larger to account for potentially missing detected systems. The value has been included in the final binary fraction calculation included in Table~\ref{tab:results}. However, as with the line-of-sight pairings, this error only applies to the spectroscopically confirmed RSG binaries and not those classified through the k-NN photometric approach. Thus, this error is again taken to be a worst-case approximation. 

\subsection{Final Binary Fraction}
To calculate a final binary fraction, the observed RSG+OB binary fraction calculated using the k-NN approach must be combined with corrections for the observational biases of our inability to detect companions fainter than B3Vs, eclipsing binaries at a single epoch, line-of-sight pairings, as well as RSGs with compact companions. Additionally, as alluded to in Section 3.1.2, calculating the overall RSG content of M31 was hampered by heavy reddening and thus I've calculated the binary fraction for all of M31 as well as just the less-reddened regions (as defined by \citealt{M3133RSGs}). The numerical results are presented as a function of metallicity in Table~\ref{tab:results} with a further discussion in the next section. Errors on the observed RSG+OB binary fraction using the k-NN approach were calculated assuming the most extreme scenarios following the given accuracy rates as given in Table~\ref{tab:kNN}. Errors for BPASS provided numbers were calculated using basic Poisson statistics. Final errors were added in quadrature after taking into account the 10\% increase in the binary fraction caused by detection limits on faint B companions.

\section{RSG Binary Fraction as a Function of Metallicity}
Examining the binary fraction of RSGs in M31 and M33 makes it possible to place constraints on any metallicity dependence. While the exact values for the metallicities of M31 and M33 vary based on measurement method (discussed further in Section 5.3), all studies point to M33 having a clear decrease in metallicity with increasing $\rho$ and M31 having little spatial metallicity dependence. Here I show that the RSG+OB binary fraction {\it also} decreases with $\rho$ in M33 but stays relatively constant throughout all of M31, suggesting a metallicity dependence on the RSG+OB binary fraction.

\subsection{A Clear Trend}
\begin{figure*}
\includegraphics[width=0.5\textwidth]{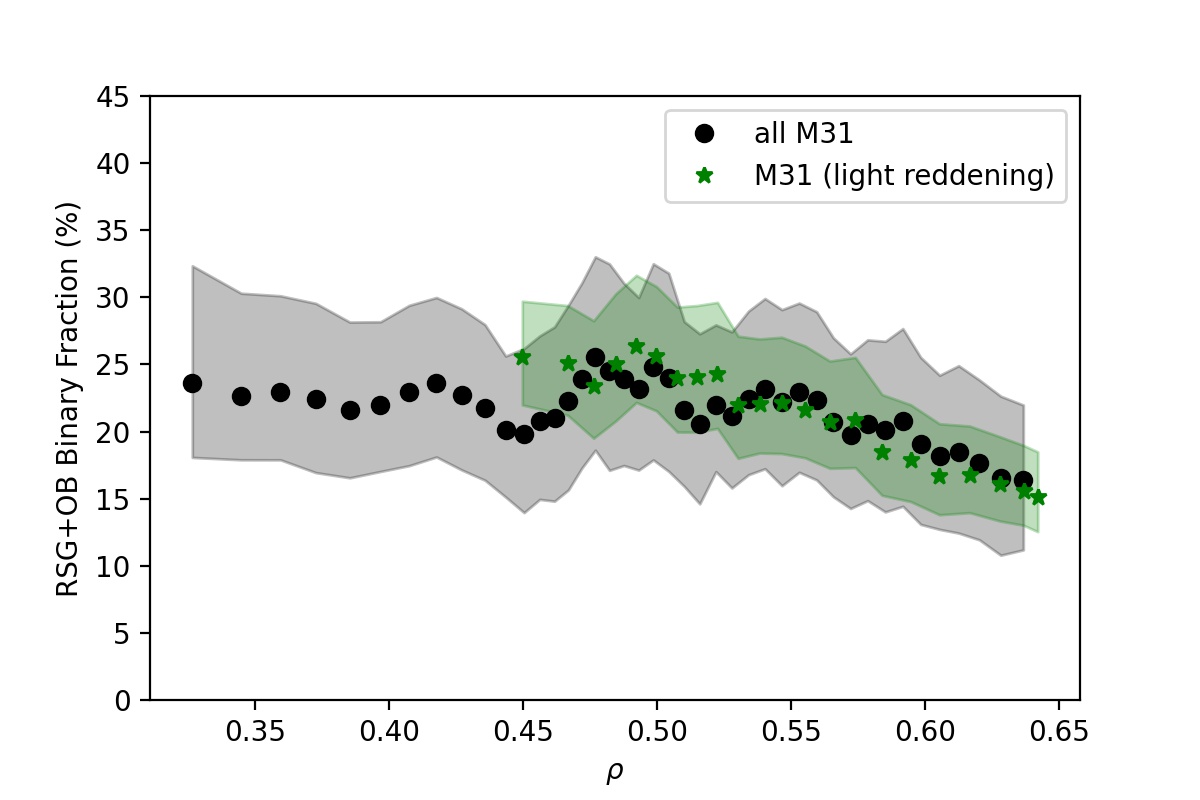}
\includegraphics[width=0.5\textwidth]{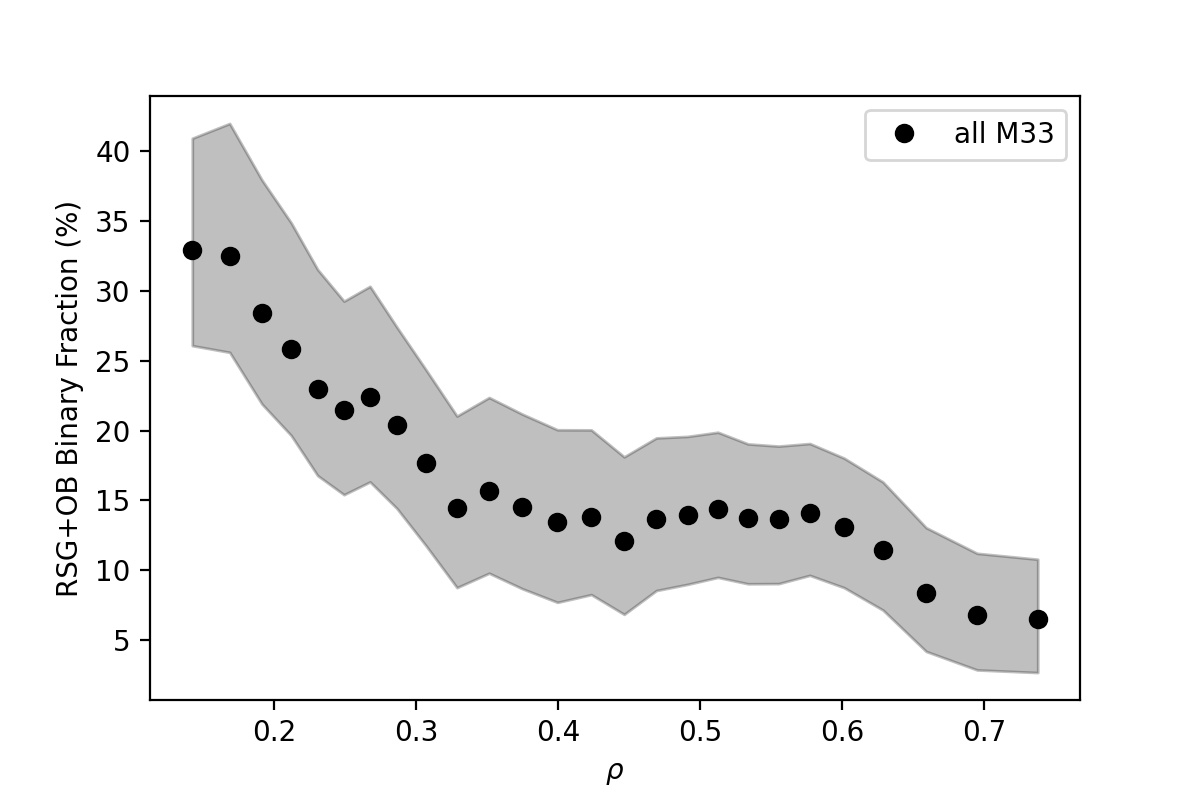}
\caption{\label{fig:binFrac} RSG+OB Binary Fraction as a function of $\rho$ (Galactocentric distance) in M31 (left) and M33 (right). The RSG+OB Binary Fraction was measured using the k-NN algorithm, as discussed in the text. Values were calculated using a moving box-car average where each point represents the average binary fraction of the closest 400 stars in $\rho$ value. Each new point is a shift of 25 stars. The strong decrease in the RSG+OB binary fraction with respect to increasing $\rho$ in M33 is due to the galaxy's strong metallicity gradient. Such a gradient does not exist in M31 and thus the binary fraction is relatively constant with increasing $\rho$.}
\end{figure*}

As seen in Table~\ref{tab:results}, there is a clear trend in both the observed RSG+OB binary fraction (as computed using the k-NN approach) as well as in the final binary fraction with increasing metallicity in M33. As the metallicity approximately doubles between the outer and inner region of M33, so does the RSG+OB binary fraction. Figure~\ref{fig:binFrac} shows the change in the RSG+OB binary fraction as a function of $\rho$ for both M31 and M33. For both galaxies, the binary fraction is plotted as a moving box-car average with each point on the plot representing the average binary fraction over the closest 400 stars in $\rho$ value. The moving average shifts by 25 stars for each new datapoint. So, taking M33 as an example, the innermost point at $\rho \sim 0.1$ and RSG+OB binary fraction at $\sim 33$\% represents the average binary fraction of the innermost 400 stars. The next point to the right represents the average RSG+OB binary fraction of the 25th - 425th innermost stars (at a $\rho\sim 0.15$ and binary fraction of $\sim 32$\%). Because each point represents 400 stars, the $\rho$ value on the x-axis stops short of the largest $\rho$ values in the dataset (in M31, $\rho$ stops at 0.75, but the stars extended to $\rho = 1$; in M33, $\rho$ stops at $\sim 0.65$ but the stars extend out to $\rho = 0.75$). Both positive and negative errors are then shown by the shaded region. 

Using this visualization method in M33, the striking metallicity dependence on the RSG+OB binary fraction becomes clear. In the innermost region of M33, where the metallicity is close to solar, the RSG+OB binary fraction is $\sim33$\%. In the outermost region where the metallicity is $0.25\times$ solar, the RSG+OB binary fraction has decreased to less than 10\%. Conversely, in M31, where there is no strong metallicity gradient, the RSG+OB binary fraction stays relatively constant at around 23\%. 

Also interesting is the presence of a small (1\%) bump in M31's RSG+OB binary fraction at $\rho \sim 0.5$. This traces the locations of the well-known star-formation ring where most of the HI, OB associations, and H$\alpha$ emission are located (see, e.g.,  \citealt{1966ApJ...144..639R,vandenbergh2000,1994AJ....108.1667D}, respectively). One criticism of the photometric/spectroscopic method I've employed is that it might be overly sensitive to line-of-sight pairings \citep{Dorda2020}. As discussed in Section 4.2, I believe these introduce a $\sim 1\%$ error, which this plot supports. If a significant fraction of the RSG+OB binaries I've detected were actually line-of-sight pairings, the RSG+OB binary fraction would be {\it much higher} in the OB forming regions (such as at $\rho \sim 0.5$ in M31) where the majority of the OB stars exist. The RSG+OB binary fraction would then be much lower in the regions further away from any OB associations. However, as shown in Figure~\ref{fig:binFrac}, this is simply not the case. The RSG+OB binary fraction in M31 is relatively constant and any small increase at $\rho \sim 0.5$ is within the errors. Thus, this lends further credence to the idea that line-of-sight pairings are not playing a large role in the detection statistics. Additionally, line-of-sight pairings due to crowding are also ruled out. If crowding were a significant issue, there would be a clear trend in the RSG+OB binary fraction in M31 with the more crowded inner regions having a higher binary fraction. While we see this general trend in M33, we can be confident it is not due to crowding both since M33 is much less crowded than M31, and such a trend is not present in the M31 data.

As discussed extensively in \citet{M3133RSGs} (in particular, see Section 5), the determination of a {\it complete} RSG sample in M31 is partially hampered by reddening within M31. Since we observe the galaxy at a significant tilt instead of face-on, some stars are redder and fainter due to us looking through extra dust. \citet{M3133RSGs} attempted to trace the regions of increased reddening by looking for RSGs with matches in the LGGS. If RSGs have matches in the LGGS, this means they were detected in $B$,$V$, and $R$, suggesting that they are lightly reddened. RSGs without matches in the LGGS were not detected in $B$ due to the large amount of dust obscuring our view and making them non-detections in the $B$ filter. Figure 9 in \citet{M3133RSGs} shows spatially the RSGs with NIR colors that also have matches in the LGGS and the resulting image matches up nicely with extinction maps of M31, notably \citet{Draine2014}. Using only the RSGs in less-reddened regions (the bright regions in Figure 9 in \citealt{M3133RSGs}), I calculated a revised RSG+OB binary fraction. While the overall percentage doesn't change much (20.4\% for all of M31 and 22.5\% for the lightly reddened regions), the uncertainties decrease by a percentage on each side due to the increased number of RSGs with LGGS colors. 

\subsection{Possible Explanations}
The results shown in Figure~\ref{fig:binFrac} and Table~\ref{tab:results} are quite striking -- there is a clear dependence of the RSG+OB binary fraction on metallicity, with higher metallicities corresponding to an increase in the binary fraction. I can first try to explain this trend by considering how the metallicity of an environment affects stellar evolution and the physical properties of RSGs. As discussed in the introduction, the metallicity of an environment has a lasting impact on the evolution of massive stars. \citet{Elias1985} first noted that as the metallicity increases, so does the average temperature of the RSGs. This was further investigated by \citet{Levesque05} by determining the physical properties of a sample of RSGs in the Magellanic Clouds and confirming that as the metallicity increases, so does the average effective temperature. Their Figure 10 shows the evolutionary tracks as a function of metallicity and the resulting shift of the Hayashi limit. As the metallicity decreases, the Hayashi limit shifts to cooler temperatures, leading to a decrease not just in the overall effective temperatures of the RSGs, but also a decrease in their maximum sustainable radii. Is this decrease in radius with metallicity significant enough that it could change the binary fraction?

As the average RSG radii increases with increasing metallicity, one might naively expect the number of RSG+OB binaries to decrease since more of them will have merged due to the increasing size of the RSG. However, this goes in the {\it opposite} direction of what we observe. To answer this question a bit more carefully, I took a look at the predications given by the BPASS v2.2.1 models. Overall, BPASS does not predict any significant change in the RSG+OB binary fraction as a function of metallicity. Instead, much like the value presented for BPASS in \citet{NeugentRSGIII}, it provides a RSG+OB binary fraction of $\sim 34-35$\% regardless of metallicity. Since BPASS {\it does} take into account the average change in RSG as a function of radius, but still does not predict a changing RSG+OB binary fraction, this leads me to believe the shift in Hayashi limit (and resulting change in physical properties) is not the main reason for the metallicity dependence.

Instead, I believe the answer lies in the formation of these stars. Little work has been done on the metallicity dependence of large populations of un-evolved OB stars as a function of metallicity. While the overall binary fraction of these stars is thought to be quite high \citep{SanaSci,Sana30Dor}, there have been no large-scale studies to show whether their binary fraction is metallicity dependent. Given that BPASS assumes a constant OB binary fraction as a function of metallicity, it then directly follows that a metallicity trend would not be revealed in the results. While there is still much work left to be done, I believe this points to a strong suggestion that the change in the RSG binary fraction with metallicity is not due to the physical properties of the RSGs changing, but rather due to metallicity changes with their parent population of OB stars either in terms of their binary fraction or the orbital separations between such systems. 

\subsection{Comments on the Metallicities of M31 and M33}
So far, I've discussed the metallicity gradient of M33 and the relatively high metallicity of M31 but I only briefly mentioned actual values in the introduction and Table~\ref{tab:results}. This is because the precise values are still open to debate and change pretty significantly based on measurement technique. There are two main methods of measuring the oxygen abundance in H\,{\sc ii} regions. The first is the ``direct" method where the the electron temperature is measured, usually from the auroral [O\,{\sc iii}] $\lambda4363$ line while the second is the ``strong-lined" method (or ``$R_{23}$") which uses the ratio of the sum of [O\,{\sc ii}] and [O\,{\sc iii}] over H$\beta$. In Table~\ref{tab:results}, I've given the values using the $R_{23}$ method from \citet{Sanders} and \citet{Magrini2007} for M31 and M33, respectively. However, care should be taken when considering these {\it exact} values as opposed to the values in a relative sense.

Based on the clear trend of decreasing RSG+OB binary fraction with $\rho$ in M33 and lack of any such trend in M31, I feel confident that this is a metallicity effect. While the exact values for the metallicity of M31 and M33 might be under some debate, there is general agreement that M33 has a strong gradient with $\rho$ while M31 does not \citep{Esteban2020, TSC2016}. Additionally, the overall LMC RSG binary fraction of $19.5^{+7.6}_{-6.7}$\% found by \citet{NeugentRSGIII} (with a metallicity somewhere between the inner and middle regions of M33) fits nicely in with the results shown in Table~\ref{tab:results}. Somewhat surprisingly, the M31 RSG binary fraction is somewhat lower than expected if all of M31 has a higher metallicity than that of the inner region of M33. But, some measurement techniques suggest that the inner regions of M33 could be higher than M31 \citep{Zaritsky94}, and thus further studies are needed. Overall, while the exact values of the metallicities for these galaxies are unknown, the trend is clear. 

\section{Summary and Next Steps}
This paper presents the first results of the RSG binary fraction as a function of metallicity using the RSG content of M31 and M33. In M33, the overall RSG binary fraction appears to range between $41.2^{+12.0}_{-7.3}$\% in the inner (higher metallicity) regions of M33 to $15.9^{+12.4}_{-1.9}$\% in the lower metallicity outer regions. In M31 where there is no metallicity gradient, the overall RSG binary fraction of the lightly reddened regions is $33.5^{+8.6}_{-5.0}$\%. These values were determined by first using the recent work of \citet{M3133RSGs} to determine a complete sample of RSGs in M31 and M33 down to a limiting $\log L/L_{\odot} \geq 4.2$ and then spectroscopically confirming candidate RSG+OB binaries that had been selected based off of their excess blue flux. Using archival LGGS and HST photometry, I then using a k-NN approach to estimate the binary fraction of RSG+OBs using the spectroscopically confirmed single and binary RSGs as a training set. The binary fraction was then corrected for observational biases (such as eclipsing systems and line-of-sight pairings) and systems the detection method may have missed (such as RSGs with companions that were either fainter than B3Vs or compact companions). The overall binary fractions for M31 and M33 are shown in Table~\ref{tab:results} and present a clear decreasing trend with metallicity. I believe this trend is due to a dependence on the binary fraction of un-evolved OB stars with metallicity, but this is still an active area of research and one that will be difficult to tackle without a targeted spectroscopic survey.

These results fit in well with previous studies of the RSG binary fraction in other environments such as in the LMC ($\log O/H +12 = 8.4$, similar to the middle region of M33; \citealt{RussellDopita}) where \citet{NeugentRSGIII} found a value of $19.5^{+7.6}_{-6.7}$\% using a similar method to the one described here. Additionally, recent radial velocity studies done by \citet{Patrick2019} and \citet{Dorda2020} placed an upper limit of the RSG binary fraction at 30\% in the LMC's 30 Doradus and a lower limit of $15\pm3$\% in both Magellanic Clouds. While such studies are more sensitive to shorter period systems, they allow for the direct detection of RSGs with compact companions. Hopefully a combination of the photometric method described here as well as the radial velocity method taken by \citet{Patrick2019} and \citet{Dorda2020} will lead to an overall better understanding of the binary fraction of RSGs.

\acknowledgements
I first thank my collaborators Emily Levesque and Phil Massey for their support and insistence that this be my first sole-authored paper. I additionally thank Meredith Durbin and Ben Williams for sharing the soon-to-be-published M33 HST PHATTER photometry and the anonymous referee for helpful suggestions that improved the paper. I additionally acknowledge that much of the work presented was done on the traditional land of the first people of Seattle, the Duwamish People past and present and honor with gratitude the land itself and the Duwamish Tribe. Observations reported here were obtained at the MMT Observatory, a joint facility of the University of Arizona and the Smithsonian Institution. This work was supported in part by a Cottrell Scholar Award from the Research Corporation for Scientific Advancement granted to Emily Levesque and by the National Science Foundation under grant AST-1612874 awarded to Phil Massey.

\newpage
\bibliographystyle{apj}
\bibliography{masterbib}

\begin{thebibliography}{}
\expandafter\ifx\csname natexlab\endcsname\relax\def\natexlab#1{#1}\fi

\bibitem[{{Bartzakos} {et~al.}(2001){Bartzakos}, {Moffat}, \&
  {Niemela}}]{Bartzakos2001}
{Bartzakos}, P., {Moffat}, A.~F.~J., \& {Niemela}, V.~S. 2001, \mnras, 324, 18

\bibitem[{{Bauer} \& {Bennett}(2000)}]{Bauer2000}
{Bauer}, W.~H., \& {Bennett}, P.~D. 2000, \pasp, 112, 31

\bibitem[{{Chatzopoulos} {et~al.}(2020){Chatzopoulos}, {Frank}, {Marcello}, \&
  {Clayton}}]{Chatzopoulos2020}
{Chatzopoulos}, E., {Frank}, J., {Marcello}, D.~C., \& {Clayton}, G.~C. 2020,
  \apj, 896, 50

\bibitem[{{Conti}(1975)}]{Conti1975}
{Conti}, P.~S. 1975, Memoires of the Societe Royale des Sciences de Liege, 9,
  193

\bibitem[{{Cox}(2000)}]{AAQ}
{Cox}, A.~N. 2000, {Allen's Astrophysical Quantities} (New York: Springer)

\bibitem[{{Dalcanton} {et~al.}(2012){Dalcanton}, {Williams}, {Lang}, {Lauer},
  {Kalirai}, {Seth}, {Dolphin}, {Rosenfield}, {Weisz}, {Bell}, {Bianchi},
  {Boyer}, {Caldwell}, {Dong}, {Dorman}, {Gilbert}, {Girardi}, {Gogarten},
  {Gordon}, {Guhathakurta}, {Hodge}, {Holtzman}, {Johnson}, {Larsen}, {Lewis},
  {Melbourne}, {Olsen}, {Rix}, {Rosema}, {Saha}, {Sarajedini}, {Skillman}, \&
  {Stanek}}]{PHAT}
{Dalcanton}, J.~J., {Williams}, B.~F., {Lang}, D., {et~al.} 2012, \apjs, 200,
  18

\bibitem[{{Devereux} {et~al.}(1994){Devereux}, {Price}, {Wells}, \&
  {Duric}}]{1994AJ....108.1667D}
{Devereux}, N.~A., {Price}, R., {Wells}, L.~A., \& {Duric}, N. 1994, \aj, 108,
  1667

\bibitem[{{Dorda} \& {Patrick}(2020)}]{Dorda2020}
{Dorda}, R., \& {Patrick}, L.~R. 2020, arXiv e-prints, arXiv:2010.15627

\bibitem[{{Dorn-Wallenstein} \& {Levesque}(2018)}]{Trevor2018}
{Dorn-Wallenstein}, T.~Z., \& {Levesque}, E.~M. 2018, \apj, 867, 125

\bibitem[{{Draine} {et~al.}(2014){Draine}, {Aniano}, {Krause}, {Groves},
  {Sandstrom}, {Braun}, {Leroy}, {Klaas}, {Linz}, {Rix}, {Schinnerer},
  {Schmiedeke}, \& {Walter}}]{Draine2014}
{Draine}, B.~T., {Aniano}, G., {Krause}, O., {et~al.} 2014, \apj, 780, 172

\bibitem[{{Drout} {et~al.}(2012){Drout}, {Massey}, \& {Meynet}}]{DroutM33RSGs}
{Drout}, M.~R., {Massey}, P., \& {Meynet}, G. 2012, \apj, 750, 97

\bibitem[{{Eldridge} {et~al.}(2017){Eldridge}, {Stanway}, {Xiao}, {McClelland
  }, {Taylor}, {Ng}, {Greis}, \& {Bray}}]{BPASS1}
{Eldridge}, J.~J., {Stanway}, E.~R., {Xiao}, L., {et~al.} 2017, \pasa, 34, e058

\bibitem[{{Elias} {et~al.}(1985){Elias}, {Frogel}, \& {Humphreys}}]{Elias1985}
{Elias}, J.~H., {Frogel}, J.~A., \& {Humphreys}, R.~M. 1985, \apjs, 57, 91

\bibitem[{{Esteban} {et~al.}(2020){Esteban}, {Bresolin}, {Garc{\'\i}a-Rojas},
  \& {Toribio San Cipriano}}]{Esteban2020}
{Esteban}, C., {Bresolin}, F., {Garc{\'\i}a-Rojas}, J., \& {Toribio San
  Cipriano}, L. 2020, \mnras, 491, 2137

\bibitem[{{Fabricant} {et~al.}(2005){Fabricant}, {Fata}, {Roll}, {Hertz},
  {Caldwell}, {Gauron}, {Geary}, {McLeod}, {Szentgyorgyi}, {Zajac}, {Kurtz},
  {Barberis}, {Bergner}, {Brown}, {Conroy}, {Eng}, {Geller}, {Goddard},
  {Honsa}, {Mueller}, {Mink}, {Ordway}, {Tokarz}, {Woods}, {Wyatt}, {Epps}, \&
  {Dell'Antonio}}]{Fabricant2005}
{Fabricant}, D., {Fata}, R., {Roll}, J., {et~al.} 2005, \pasp, 117, 1411

\bibitem[{{Foellmi} {et~al.}(2003{\natexlab{a}}){Foellmi}, {Moffat}, \&
  {Guerrero}}]{FoellmiSMC}
{Foellmi}, C., {Moffat}, A.~F.~J., \& {Guerrero}, M.~A. 2003{\natexlab{a}},
  \mnras, 338, 360

\bibitem[{{Foellmi} {et~al.}(2003{\natexlab{b}}){Foellmi}, {Moffat}, \&
  {Guerrero}}]{Foellmi2003}
---. 2003{\natexlab{b}}, \mnras, 338, 1025

\bibitem[{{Gaia Collaboration} {et~al.}(2018){Gaia Collaboration}, {Helmi},
  {van Leeuwen}, {McMillan}, {Massari}, {Antoja}, {Robin}, {Lindegren},
  {Bastian}, {Arenou}, {Babusiaux}, {Biermann}, {Breddels}, {Hobbs}, {Jordi},
  {Pancino}, {Reyl{\'e}}, {Veljanoski}, {Brown}, {Vallenari}, {Prusti}, {de
  Bruijne}, {Bailer-Jones}, {Evans}, {Eyer}, {Jansen}, {Klioner}, {Lammers},
  {Luri}, {Mignard}, {Panem}, {Pourbaix}, {Randich}, {Sartoretti}, {Siddiqui},
  {Soubiran}, {Walton}, {Cropper}, {Drimmel}, {Katz}, {Lattanzi}, {Bakker},
  {Cacciari}, {Casta{\~n}eda}, {Chaoul}, {Cheek}, {De Angeli}, {Fabricius},
  {Guerra}, {Holl}, {Masana}, {Messineo}, {Mowlavi}, {Nienartowicz}, {Panuzzo},
  {Portell}, {Riello}, {Seabroke}, {Tanga}, {Th{\'e}venin}, {Gracia-Abril},
  {Comoretto}, {Garcia-Reinaldos}, {Teyssier}, {Altmann}, {Andrae}, {Audard},
  {Bellas-Velidis}, {Benson}, {Berthier}, {Blomme}, {Burgess}, {Busso},
  {Carry}, {Cellino}, {Clementini}, {Clotet}, {Creevey}, {Davidson}, {De
  Ridder}, {Delchambre}, {Dell'Oro}, {Ducourant},
  {Fern{\'a}ndez-Hern{\'a}ndez}, {Fouesneau}, {Fr{\'e}mat}, {Galluccio},
  {Garc{\'\i}a-Torres}, {Gonz{\'a}lez-N{\'u}{\~n}ez}, {Gonz{\'a}lez-Vidal},
  {Gosset}, {Guy}, {Halbwachs}, {Hambly}, {Harrison}, {Hern{\'a}ndez},
  {Hestroffer}, {Hodgkin}, {Hutton}, {Jasniewicz}, {Jean-Antoine-Piccolo},
  {Jordan}, {Korn}, {Krone-Martins}, {Lanzafame}, {Lebzelter}, {L{\"o}ffler},
  {Manteiga}, {Marrese}, {Mart{\'\i}n-Fleitas}, {Moitinho}, {Mora}, {Muinonen},
  {Osinde}, {Pauwels}, {Petit}, {Recio-Blanco}, {Richards}, {Rimoldini},
  {Sarro}, {Siopis}, {Smith}, {Sozzetti}, {S{\"u}veges}, {Torra}, {van Reeven},
  {Abbas}, {Abreu Aramburu}, {Accart}, {Aerts}, {Altavilla}, {{\'A}lvarez},
  {Alvarez}, {Alves}, {Anderson}, {Andrei}, {Anglada Varela}, {Antiche},
  {Arcay}, {Astraatmadja}, {Bach}, {Baker}, {Balaguer-N{\'u}{\~n}ez}, {Balm},
  {Barache}, {Barata}, {Barbato}, {Barblan}, {Barklem}, {Barrado}, {Barros},
  {Barstow}, {Bartholom{\'e} Mu{\~n}oz}, {Bassilana}, {Becciani}, {Bellazzini},
  {Berihuete}, {Bertone}, {Bianchi}, {Bienaym{\'e}}, {Blanco-Cuaresma}, {Boch},
  {Boeche}, {Bombrun}, {Borrachero}, {Bossini}, {Bouquillon}, {Bourda},
  {Bragaglia}, {Bramante}, {Bressan}, {Brouillet}, {Br{\"u}semeister},
  {Brugaletta}, {Bucciarelli}, {Burlacu}, {Busonero}, {Butkevich}, {Buzzi},
  {Caffau}, {Cancelliere}, {Cannizzaro}, {Cantat-Gaudin}, {Carballo},
  {Carlucci}, {Carrasco}, {Casamiquela}, {Castellani}, {Castro-Ginard},
  {Charlot}, {Chemin}, {Chiavassa}, {Cocozza}, {Costigan}, {Cowell}, {Crifo},
  {Crosta}, {Crowley}, {Cuypers}, {Dafonte}, {Damerdji}, {Dapergolas}, {David},
  {David}, {de Laverny}, {De Luise}, {De March}, {de Martino}, {de Souza}, {de
  Torres}, {Debosscher}, {del Pozo}, {Delbo}, {Delgado}, {Delgado}, {Di
  Matteo}, {Diakite}, {Diener}, {Distefano}, {Dolding}, {Drazinos},
  {Dur{\'a}n}, {Edvardsson}, {Enke}, {Eriksson}, {Esquej}, {Eynard Bontemps},
  {Fabre}, {Fabrizio}, {Faigler}, {Falc{\~a}o}, {Farr{\`a}s Casas}, {Federici},
  {Fedorets}, {Fernique}, {Figueras}, {Filippi}, {Findeisen}, {Fonti},
  {Fraile}, {Fraser}, {Fr{\'e}zouls}, {Gai}, {Galleti}, {Garabato},
  {Garc{\'\i}a-Sedano}, {Garofalo}, {Garralda}, {Gavel}, {Gavras}, {Gerssen},
  {Geyer}, {Giacobbe}, {Gilmore}, {Girona}, {Giuffrida}, {Glass}, {Gomes},
  {Granvik}, {Gueguen}, {Guerrier}, {Guiraud}, {Guti{\'e}rrez-S{\'a}nchez},
  {Hofmann}, {Holland}, {Huckle}, {Hypki}, {Icardi}, {Jan{\ss}en}, {Jevardat de
  Fombelle}, {Jonker}, {Juh{\'a}sz}, {Julbe}, {Karampelas}, {Kewley}, {Klar},
  {Kochoska}, {Kohley}, {Kolenberg}, {Kontizas}, {Kontizas}, {Koposov},
  {Kordopatis}, {Kostrzewa-Rutkowska}, {Koubsky}, {Lambert}, {Lanza}, {Lasne},
  {Lavigne}, {Le Fustec}, {Le Poncin-Lafitte}, {Lebreton}, {Leccia}, {Leclerc},
  {Lecoeur-Taibi}, {Lenhardt}, {Leroux}, {Liao}, {Licata}, {Lindstr{\o}m},
  {Lister}, {Livanou}, {Lobel}, {L{\'o}pez}, {Managau}, {Mann}, {Mantelet},
  {Marchal}, {Marchant}, {Marconi}, {Marinoni}, {Marschalk{\'o}}, {Marshall},
  {Martino}, {Marton}, {Mary}, {Matijevi{\v{c}}}, {Mazeh}, {Messina},
  {Michalik}, {Millar}, {Molina}, {Molinaro}, {Moln{\'a}r}, {Montegriffo},
  {Mor}, {Morbidelli}, {Morel}, {Morris}, {Mulone}, {Muraveva}, {Musella},
  {Nelemans}, {Nicastro}, {Noval}, {O'Mullane}, {Ord{\'e}novic},
  {Ord{\'o}{\~n}ez-Blanco}, {Osborne}, {Pagani}, {Pagano}, {Pailler},
  {Palacin}, {Palaversa}, {Panahi}, {Pawlak}, {Piersimoni}, {Pineau}, {Plachy},
  {Plum}, {Poggio}, {Poujoulet}, {Pr{\v{s}}a}, {Pulone}, {Racero}, {Ragaini},
  {Rambaux}, {Ramos-Lerate}, {Regibo}, {Riclet}, {Ripepi}, {Riva}, {Rivard},
  {Rixon}, {Roegiers}, {Roelens}, {Romero-G{\'o}mez}, {Rowell}, {Royer},
  {Ruiz-Dern}, {Sadowski}, {Sagrist{\`a} Sell{\'e}s}, {Sahlmann}, {Salgado},
  {Salguero}, {Sanna}, {Santana-Ros}, {Sarasso}, {Savietto}, {Schultheis},
  {Sciacca}, {Segol}, {Segovia}, {S{\'e}gransan}, {Shih}, {Siltala}, {Silva},
  {Smart}, {Smith}, {Solano}, {Solitro}, {Sordo}, {Soria Nieto}, {Souchay},
  {Spagna}, {Spoto}, {Stampa}, {Steele}, {Steidelm{\"u}ller}, {Stephenson},
  {Stoev}, {Suess}, {Surdej}, {Szabados}, {Szegedi-Elek}, {Tapiador}, {Taris},
  {Tauran}, {Taylor}, {Teixeira}, {Terrett}, {Teyssand ier}, {Thuillot},
  {Titarenko}, {Torra Clotet}, {Turon}, {Ulla}, {Utrilla}, {Uzzi}, {Vaillant},
  {Valentini}, {Valette}, {van Elteren}, {Van Hemelryck}, {van Leeuwen},
  {Vaschetto}, {Vecchiato}, {Viala}, {Vicente}, {Vogt}, {von Essen}, {Voss},
  {Votruba}, {Voutsinas}, {Walmsley}, {Weiler}, {Wertz}, {Wevems},
  {Wyrzykowski}, {Yoldas}, {{\v{Z}}erjal}, {Ziaeepour}, {Zorec}, {Zschocke},
  {Zucker}, {Zurbach}, \& {Zwitter}}]{Gaia}
{Gaia Collaboration}, {Helmi}, A., {van Leeuwen}, F., {et~al.} 2018, \aap, 616,
  A12

\bibitem[{{Gonz{\'a}lez-Fern{\'a}ndez}
  {et~al.}(2015){Gonz{\'a}lez-Fern{\'a}ndez}, {Dorda}, {Negueruela}, \&
  {Marco}}]{GF2015}
{Gonz{\'a}lez-Fern{\'a}ndez}, C., {Dorda}, R., {Negueruela}, I., \& {Marco}, A.
  2015, \aap, 578, A3

\bibitem[{{Hinkle} {et~al.}(2020){Hinkle}, {Lebzelter}, {Fekel}, {Straniero},
  {Joyce}, {Prato}, {Karnath}, \& {Habel}}]{Hinkle2020}
{Hinkle}, K.~H., {Lebzelter}, T., {Fekel}, F.~C., {et~al.} 2020, arXiv
  e-prints, arXiv:2010.01081

\bibitem[{{Kippenhahn} \& {Weigert}(1967)}]{Kippenhahn1967}
{Kippenhahn}, R., \& {Weigert}, A. 1967, \zap, 65, 251

\bibitem[{{Lanz} \& {Hubeny}(2007)}]{BSTARS06}
{Lanz}, T., \& {Hubeny}, I. 2007, \apjs, 169, 83

\bibitem[{{Levesque} {et~al.}(2005){Levesque}, {Massey}, {Olsen}, {Plez},
  {Josselin}, {Maeder}, \& {Meynet}}]{Levesque05}
{Levesque}, E.~M., {Massey}, P., {Olsen}, K.~A.~G., {et~al.} 2005, \apj, 628,
  973

\bibitem[{{Magrini} {et~al.}(2007){Magrini}, {V{\'\i}lchez}, {Mampaso},
  {Corradi}, \& {Leisy}}]{Magrini2007}
{Magrini}, L., {V{\'\i}lchez}, J.~M., {Mampaso}, A., {Corradi}, R.~L.~M., \&
  {Leisy}, P. 2007, \aap, 470, 865

\bibitem[{{Martin} \& {GALEX Team}(2005)}]{Martin05}
{Martin}, C., \& {GALEX Team}. 2005, in Multiwavelength Mapping of Galaxy
  Formation and Evolution, ed. A.~{Renzini} \& R.~{Bender}, 197

\bibitem[{{Massey}(1998)}]{Massey98}
{Massey}, P. 1998, \apj, 501, 153

\bibitem[{{Massey} \& {Evans}(2016)}]{MasseyEvans16}
{Massey}, P., \& {Evans}, K.~A. 2016, \apj, 826, 224

\bibitem[{{Massey} {et~al.}(2007){Massey}, {McNeill}, {Olsen}, {Hodge},
  {Blaha}, {Jacoby}, {Smith}, \& {Strong}}]{LGGSIII}
{Massey}, P., {McNeill}, R.~T., {Olsen}, K.~A.~G., {et~al.} 2007, \aj, 134,
  2474

\bibitem[{{Massey} {et~al.}(2019){Massey}, {Neugent}, \&
  {Levesque}}]{Massey2019}
{Massey}, P., {Neugent}, K.~F., \& {Levesque}, E.~M. 2019, \aj, 157, 227

\bibitem[{{Massey} {et~al.}(2020){Massey}, {Neugent}, {Levesque}, {Drout}, \&
  {Courteau}}]{M3133RSGs}
{Massey}, P., {Neugent}, K.~N., {Levesque}, E.~M., {Drout}, M.~R., \&
  {Courteau}, S. 2020, \aj, submitted

\bibitem[{{Massey} {et~al.}(2006){Massey}, {Olsen}, {Hodge}, {Strong},
  {Jacoby}, {Schlingman}, \& {Smith}}]{LGGS}
{Massey}, P., {Olsen}, K.~A.~G., {Hodge}, P.~W., {et~al.} 2006, \aj, 131, 2478

\bibitem[{{Moe} \& {Di Stefano}(2017)}]{Moe2017}
{Moe}, M., \& {Di Stefano}, R. 2017, \apjs, 230, 15

\bibitem[{{Moe} {et~al.}(2019){Moe}, {Kratter}, \& {Badenes}}]{Moe2019}
{Moe}, M., {Kratter}, K.~M., \& {Badenes}, C. 2019, \apj, 875, 61

\bibitem[{{Morrissey} {et~al.}(2007){Morrissey}, {Conrow}, {Barlow}, {Small},
  {Seibert}, {Wyder}, {Budav{\'a}ri}, {Arnouts}, {Friedman}, {Forster},
  {Martin}, {Neff}, {Schiminovich}, {Bianchi}, {Donas}, {Heckman}, {Lee},
  {Madore}, {Milliard}, {Rich}, {Szalay}, {Welsh}, \& {Yi}}]{Morrissey07}
{Morrissey}, P., {Conrow}, T., {Barlow}, T.~A., {et~al.} 2007, \apjs, 173, 682

\bibitem[{{Neugent} {et~al.}(2018){Neugent}, {Levesque}, \&
  {Massey}}]{NeugentRSG}
{Neugent}, K.~F., {Levesque}, E.~M., \& {Massey}, P. 2018, \aj, 156, 225

\bibitem[{{Neugent} {et~al.}(2019){Neugent}, {Levesque}, {Massey}, \&
  {Morrell}}]{NeugentRSGII}
{Neugent}, K.~F., {Levesque}, E.~M., {Massey}, P., \& {Morrell}, N.~I. 2019,
  \apj, 875, 124

\bibitem[{{Neugent} {et~al.}(2020{\natexlab{a}}){Neugent}, {Levesque},
  {Massey}, {Morrell}, \& {Drout}}]{NeugentRSGIII}
{Neugent}, K.~F., {Levesque}, E.~M., {Massey}, P., {Morrell}, N.~I., \&
  {Drout}, M.~R. 2020{\natexlab{a}}, arXiv e-prints, arXiv:2007.15852

\bibitem[{{Neugent} \& {Massey}(2011)}]{M33WRs}
{Neugent}, K.~F., \& {Massey}, P. 2011, \apj, 733, 123

\bibitem[{{Neugent} \& {Massey}(2014)}]{WRbins}
---. 2014, \apj, 789, 10

\bibitem[{{Neugent} {et~al.}(2012){Neugent}, {Massey}, \& {Georgy}}]{M31WRs}
{Neugent}, K.~F., {Massey}, P., \& {Georgy}, C. 2012, \apj, 759, 11

\bibitem[{{Neugent} {et~al.}(2020{\natexlab{b}}){Neugent}, {Massey}, {Georgy},
  {Drout}, {Mommert}, {Levesque}, {Meynet}, \& {Ekstr{\"o}m}}]{UKIRT}
{Neugent}, K.~F., {Massey}, P., {Georgy}, C., {et~al.} 2020{\natexlab{b}},
  \apj, 889, 44

\bibitem[{{Patrick} {et~al.}(2019){Patrick}, {Lennon}, {Britavskiy}, {Evans},
  {Sana}, {Taylor}, {Herrero}, {Almeida}, {Clark}, {Gieles}, {Langer},
  {Schneider}, \& {van Loon}}]{Patrick2019}
{Patrick}, L.~R., {Lennon}, D.~J., {Britavskiy}, N., {et~al.} 2019, \aap, 624,
  A129

\bibitem[{{Plez} {et~al.}(1992){Plez}, {Brett}, \& {Nordlund}}]{MARCS}
{Plez}, B., {Brett}, J.~M., \& {Nordlund}, A. 1992, \aap, 256, 551

\bibitem[{{Roberts}(1966)}]{1966ApJ...144..639R}
{Roberts}, M.~S. 1966, \apj, 144, 639

\bibitem[{{Russell} \& {Dopita}(1990)}]{RussellDopita}
{Russell}, S.~C., \& {Dopita}, M.~A. 1990, \apjs, 74, 93

\bibitem[{{Sana} {et~al.}(2012){Sana}, {de Mink}, {de Koter}, {Langer},
  {Evans}, {Gieles}, {Gosset}, {Izzard}, {Le Bouquin}, \&
  {Schneider}}]{SanaSci}
{Sana}, H., {de Mink}, S.~E., {de Koter}, A., {et~al.} 2012, Science, 337, 444

\bibitem[{{Sana} {et~al.}(2013){Sana}, {de Koter}, {de Mink}, {Dunstall},
  {Evans}, {H{\'e}nault-Brunet}, {Ma{\'\i}z Apell{\'a}niz},
  {Ram{\'\i}rez-Agudelo}, {Taylor}, {Walborn}, {Clark}, {Crowther}, {Herrero},
  {Gieles}, {Langer}, {Lennon}, \& {Vink}}]{Sana30Dor}
{Sana}, H., {de Koter}, A., {de Mink}, S.~E., {et~al.} 2013, \aap, 550, A107

\bibitem[{{Sanders} {et~al.}(2012){Sanders}, {Caldwell}, {McDowell}, \&
  {Harding}}]{Sanders}
{Sanders}, N.~E., {Caldwell}, N., {McDowell}, J., \& {Harding}, P. 2012, \apj,
  758, 133

\bibitem[{{Simons} {et~al.}(2014){Simons}, {Thilker}, {Bianchi}, \&
  {Wyder}}]{Simons14}
{Simons}, R., {Thilker}, D., {Bianchi}, L., \& {Wyder}, T. 2014, Advances in
  Space Research, 53, 939

\bibitem[{{Skrutskie} {et~al.}(2006){Skrutskie}, {Cutri}, {Stiening},
  {Weinberg}, {Schneider}, {Carpenter}, {Beichman}, {Capps}, {Chester},
  {Elias}, {Huchra}, {Liebert}, {Lonsdale}, {Monet}, {Price}, {Seitzer},
  {Jarrett}, {Kirkpatrick}, {Gizis}, {Howard}, {Evans}, {Fowler}, {Fullmer},
  {Hurt}, {Light}, {Kopan}, {Marsh}, {McCallon}, {Tam}, {Van Dyk}, \&
  {Wheelock}}]{2MASS}
{Skrutskie}, M.~F., {Cutri}, R.~M., {Stiening}, R., {et~al.} 2006, \aj, 131,
  1163

\bibitem[{{Stanway} \& {Eldridge}(2018)}]{BPASS2}
{Stanway}, E.~R., \& {Eldridge}, J.~J. 2018, \mnras, 479, 75

\bibitem[{{Toribio San Cipriano} {et~al.}(2016){Toribio San Cipriano},
  {Garc{\'\i}a-Rojas}, {Esteban}, {Bresolin}, \& {Peimbert}}]{TSC2016}
{Toribio San Cipriano}, L., {Garc{\'\i}a-Rojas}, J., {Esteban}, C., {Bresolin},
  F., \& {Peimbert}, M. 2016, \mnras, 458, 1866

\bibitem[{{van den Bergh}(2000)}]{vandenbergh2000}
{van den Bergh}, S. 2000, {The Galaxies of the Local Group} (Cambridge:
  Cambridge Univ.\ Press)

\bibitem[{{Wheeler} {et~al.}(2017){Wheeler}, {Nance}, {Diaz}, {Smith},
  {Hickey}, {Zhou}, {Koutoulaki}, {Sullivan}, \& {Fowler}}]{Wheeler17}
{Wheeler}, J.~C., {Nance}, S., {Diaz}, M., {et~al.} 2017, \mnras, 465, 2654

\bibitem[{{Yang} {et~al.}(2019){Yang}, {Bonanos}, {Jiang}, {Gao}, {Gavras},
  {Maravelias}, {Ren}, {Wang}, {Xue}, {Tramper}, {Spetsieri}, \&
  {Pouliasis}}]{Yang2019}
{Yang}, M., {Bonanos}, A.~Z., {Jiang}, B.-W., {et~al.} 2019, \aap, 629, A91

\bibitem[{{Yoon} {et~al.}(2017){Yoon}, {Dessart}, \& {Clocchiatti}}]{Yoon2017}
{Yoon}, S.-C., {Dessart}, L., \& {Clocchiatti}, A. 2017, \apj, 840, 10

\bibitem[{{Zaritsky} {et~al.}(1994){Zaritsky}, {Kennicutt}, \&
  {Huchra}}]{Zaritsky94}
{Zaritsky}, D., {Kennicutt}, Robert~C., J., \& {Huchra}, J.~P. 1994, \apj, 420,
  87

\end{thebibliography}

\end{document}